\documentclass[traditabstract]{aa}

\usepackage{setspace}
\usepackage{graphicx}
\usepackage{natbib}
\usepackage{txfonts}
\bibpunct{(}{)}{;}{a}{}{,} 

\begin{document}

\def\u{\rlap{:}}
\def\uu{\rlap{::}}
\def\m{\rlap{$^a$}}
\def\x{\hspace{3mm}}
\def\ms#1{#1]}

\title{A Herschel study of NGC 650\thanks{Herschel is an ESA space observatory with science instruments provided by European-led Principal Investigator consortia and with important participation from NASA.}}

\author{P.A.M. van Hoof\inst{1}\thanks{email: p.vanhoof@oma.be}
  \and
  G.C. Van de Steene\inst{1}
  \and
  K.M. Exter\inst{2}
  \and
  M.J. Barlow\inst{3}
  \and
  T.~Ueta\inst{4}
  \and
  M.A.T.~Groenewegen\inst{1}
  \and
  W.K.~Gear\inst{5}
  \and
  H.L.~Gomez\inst{5}
  \and
  P.C.~Hargrave\inst{5}
  \and
  R.J.~Ivison\inst{6}
  \and
  S.J.~Leeks\inst{7}
  \and
  T.L.~Lim\inst{7}
  \and
  G.~Olofsson\inst{8}
  \and
  E.T.~Polehampton\inst{7,9}
  \and
  B.M.~Swinyard\inst{7}
  \and
  H.~Van~Winckel\inst{2}
  \and
  C.~Waelkens\inst{2}
  \and
  R.~Wesson\inst{3}
}

\authorrunning{van Hoof et al.}

\institute{Royal Observatory of Belgium, Ringlaan 3, B-1180 Brussels, Bel\-gium
  \and
  Instituut voor Sterrenkunde, Katholieke Universiteit Leuven, Ce\-les\-tij\-nenlaan 200 D, B-3001 Leuven, Belgium
  \and
  Dept of Physics \& Astronomy, University College London, Gower St, London WC1E 6BT, UK 
  \and
  Dept. of Physics and Astronomy, University of Denver, Mail Stop 6900, Denver, CO 80208, USA 
  \and
  School of Physics and Astronomy, Cardiff University, 5 The Parade, Cardiff, Wales CF24 3YB, UK
  \and
  UK Astronomy Technology Centre, Royal Observatory Edinburgh, Blackford Hill, Edinburgh EH9 3HJ, UK
  \and
  Space Science and Technology Department, Rutherford Appleton Laboratory, Oxfordshire, OX11 0QX, UK  
  \and
  Dept.\ of Astronomy, Stockholm University, AlbaNova University Center, Roslagstullsbacken 21, 10691 Stockholm, Sweden 
  \and
  Department of Physics, University of Lethbridge, Lethbridge, Alberta, T1J 1B1, Canada 
}

\date{Received; accepted}

\abstract{As part of the Herschel Guaranteed Time Key Project MESS (Mass loss
  of Evolved StarS) we have imaged a sample of planetary nebulae. In this
  paper we present the PACS and SPIRE images of the classical bipolar
  planetary nebula NGC 650. We used these images to derive a temperature map
  of the dust. We also constructed a photoionization and dust radiative
  transfer model using the spectral synthesis code Cloudy. To constrain this
  model, we used the PACS and SPIRE fluxes and combined these with hitherto
  unpublished IUE and Spitzer IRS spectra as well as various other data from
  the literature. The temperature map combined with the photoionization model
  were used to study various aspects of the central star, the nebula, and in
  particular the dust grains in the nebula. The central star parameters are
  determined to be $T_{\rm eff} = 208$~kK and $L = 261$~$L_{\odot}$ assuming a
  distance of 1200~pc. The stellar temperature is much higher than previously
  published values. We confirm that the nebula is carbon-rich with a C/O ratio
  of 2.1. The nebular abundances are typical for a type IIa planetary nebula.
  With the photoionization model we determined that the grains in the ionized
  nebula are large (assuming single-sized grains, they would have a radius of
  0.15~$\mu$m). Most likely these large grains were inherited from the
  asymptotic giant branch phase. The PACS 70/160~$\mu$m temperature map shows
  evidence for two radiation components heating the grains. The first
  component is direct emission from the central star, while the second
  component is diffuse emission from the ionized gas (mainly Ly$\alpha$). We
  show that previous suggestions that there is a photo-dissociation region
  surrounding the ionized region are incorrect. The neutral material resides
  in dense clumps inside the ionized region. These may also harbor
  stochastically heated very small grains in addition to the large grains.}

\keywords{planetary nebulae: individual: NGC 650 --  
  circumstellar matter --
  dust, extinction --
  Infrared: ISM --
  ISM: molecules
}

\maketitle

\section{Introduction}
\label{intro}

As part of the Herschel Guaranteed Time Key Project MESS (Mass loss of Evolved
StarS) (PI Martin Groenewegen) we have imaged a sample of planetary nebulae
(PNe) with the PACS \citep{Poglitsch10} and SPIRE \citep{Griffin10}
instruments on board the Herschel satellite \citep{Pilbratt10}. A detailed
description of the program can be found in \citet{Groenewegen11}. The main
aims of the MESS program are twofold, namely (1) to study the structure of the
circumstellar envelope and time evolution of the mass-loss rate, and (2) to
study molecular and solid-state features in the spectra of a representative
sample of low, intermediate, and high-mass post-main sequence objects. An
overview of the Herschel observations for PNe can be found in
\citet{vanHoof12}. In this article we will present the Herschel observations
of the PN NGC~650.

NGC~650 (M~76, Little Dumbbell Nebula) is a large ($\sim$300\arcsec,
\citealp{Balick92}) bipolar planetary nebula (PN) of the "late butterfly" type
\citep{Balick87}. It is a prototypical example of a bipolar PN. The generally
accepted model for the formation of such a PN is the Generalized Interacting
Stellar Winds model (GISW, \citealp{Balick87}) in which the progenitor has
lost a considerable amount of mass whilst on the asymptotic giant branch
(AGB), culminating in the superwind. The GISW model depends on the degree of
equator to pole density contrast in the AGB wind to produce a wide variety of
shapes, which superficially may appear quite different. One reason for
studying NGC~650 is that it has historically been linked to the Ring nebula
(NGC~6720, \citealp{Minkowski60}), which is also part of the MESS program
\citep{vanHoof12}. NGC~6720 is an evolved, oxygen-rich bipolar nebula seen
nearly pole-on \citep{OD07,vH10}, while NGC~650 has been claimed to be carbon
rich \citep{KH96}. Both nebulae have a highly evolved central star and do
exhibit small-scale density variations and irregularities in the nebula on a
larger scale. In this paper we investigate the dust properties of NGC~650
using Cloudy modeling as we did for the Ring nebula in \citet{vH10}.

NGC~650 is a rather well studied nebula. The optical and mid-IR (mid-infrared)
properties of NGC~650 have been studied by \citet{RL08}, the far-IR Spitzer
emission by \citet{Ue06}, the kinematical properties by \citet{Bryce96}, and
the H$_2$ emission by \citet{ML13}, and references therein.

The nebular structure in the optical consists of a bright rectangular core of
100\arcsec\ $\times$ 40\arcsec\ (the long side perpendicular to the bipolar
axis) and a pair of fainter lobes extending $\sim$90\arcsec\ and
$\sim$150\arcsec\ from the central star \citep{Balick92}. The optical low
excitation intensity ratio maps show evidence for a complex and clumpy
structure throughout the source, located in the central bar-like feature
(where it appears to be particularly prominent), the semicircular extended
lobes, as well as the weaker emission beyond these lobes \citep{RL08}.

The most recent kinematical study has demonstrated that the core is an
inclined torus and the lobes are blown bubbles expanding in the polar
directions \citep{Bryce96}. The torus is likely to take the form of a tilted
`napkin-ring' type structure inclined at about 75\degr\ to the line of sight
such that the NW lobes extends toward the observer and the SE lobe extends
away. This torus is expanding at a velocity of $\sim$ 43 km\,s$^{-1}$, while
the bright semicircular lobes are observed to expand at $\sim$
60~km\,s$^{-1}$, and the fainter outer shell appears to have $v_{\rm exp}
\sim$ 5~km\,s$^{-1}$.

The mid-IR emission at 5.8~and~8~$\mu$m is likely composed of polycyclic
aromatic hydrocarbon (PAH) emission bands and is associated with the nebular
photo-dissociation region (PDR) \citep{RL08}. \citet{RL08} suggest that much
of the fainter emission outside the bipolar lobes may arise as a result of
leakage of radiation from the clumpy interior lobes.

The far-IR surface brightness distribution shows two emission peaks at all
three MIPS bands \citep{Ue06}. These peaks represent the limb-brightened edges
of a nearly edge-on, optically thin dusty torus. The 70 and 160~$\mu$m
emission comes from low-temperature dust ($\sim$30~K, see
Sect.~\ref{other:data}) in the remnant AGB wind shell. The 24 $\mu$m emission
is in part due to the [O\,{\sc iv}] line at 25.9 $\mu$m arising from the
highly ionized part of the inner torus, which has been engulfed by the
ionization front, and the ionized matter that has advected off the torus and
filled the inner cavity of the torus. However, part of the 24 $\mu$m emission
also comes from dust grains. This is discussed further in Sect.~\ref{heating}.

H$_2$ emission arises from clumps embedded within the ionized torus
\citep{ML13}. The H$_2$ observations resolve the torus into a series of
disconnected knots and filaments. Some knots are also detected in the inner
regions of the bipolar lobes. The H$_2$ emission is not constrained to the
waist only. No CO emission has been detected in NGC~650 \citep{HH89,Hu96} in
the regions where the H$_2$ emission is strongest.

This PN has a hydrogen deficient central star of type PG 1159 (E) with
estimated parameters $T_{\rm eff} = 140$~kK, $\log g = 7.0$ and $M_{\rm core}
= 0.60$~$M_\odot$ \citep{NS95}. These values were derived by comparison to the
very similar optical spectrum of PG 1159 itself, which was extensively studied
by \citet{We91}. Note however that the NGC 650 spectrum used by \citet{NS95}
is of fairly low quality (see their Fig.~7) so that it is hard to assess the
accuracy of these parameters. See also the discussion in
Sect.~\ref{star:temp}. The position of the central star is
$01^h\,42^m\,19\fs948$ $+51\degr\,34\arcmin\,31\farcs15$ (J2000).

The distance to the PN is 1200~pc \citep{KP98}. This value is a so-called
gravity distance. It is derived by comparing the stellar atmosphere model
combined with the $\log g$ value (which gives the absolute luminosity of the
star) with the observed V band magnitude after correcting for the extinction.
Given the uncertainty in the central star parameters discussed above and the
non-standard extinction in the nebula (see Sect.~\ref{dered}) this value
should probably be considered uncertain.

In Sect.~\ref{data:red} we discuss the reduction of the PACS and SPIRE data
and present the images. In Sect.~\ref{model} we discuss the photoionization
and dust radiative transfer model of NGC~650 that we created as well as the
data that we collected to constrain this model. In Sect.~\ref{discussion} we
will give a full discussion of our results and finally in
Sect.~\ref{conclusions} we will present our main conclusions.

\section{The Herschel Data}
\label{data:red}

\begin{figure}
\begin{center}
\includegraphics[width=\columnwidth]{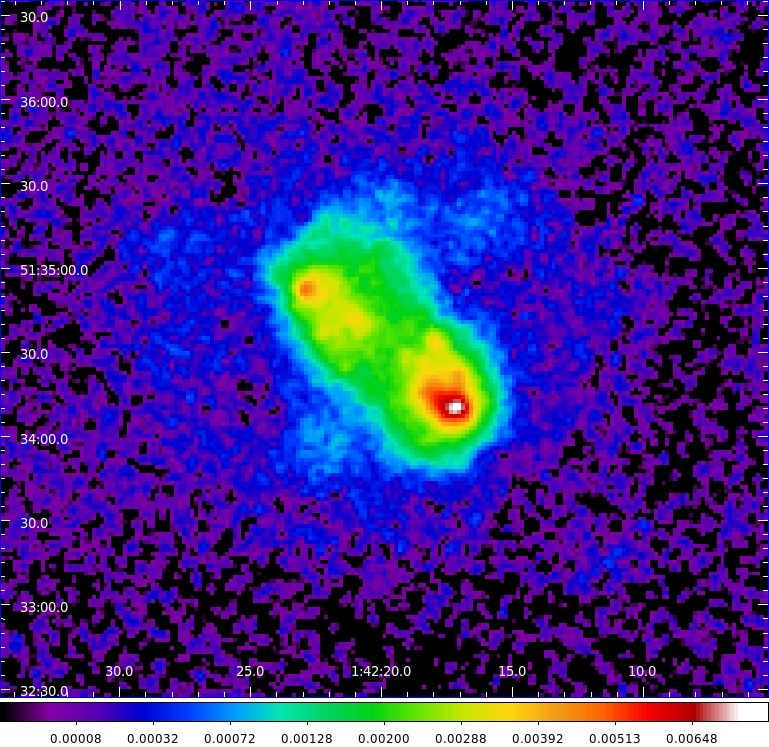} 
\caption{The PACS 70~$\mu$m image of NGC~650 shown on a sqrt stretch. The
  color bar shows the flux density normalized in Jy/pixel.}
\label{fig70}
\end{center}
\end{figure}

\begin{figure}
\begin{center}
\includegraphics[width=\columnwidth]{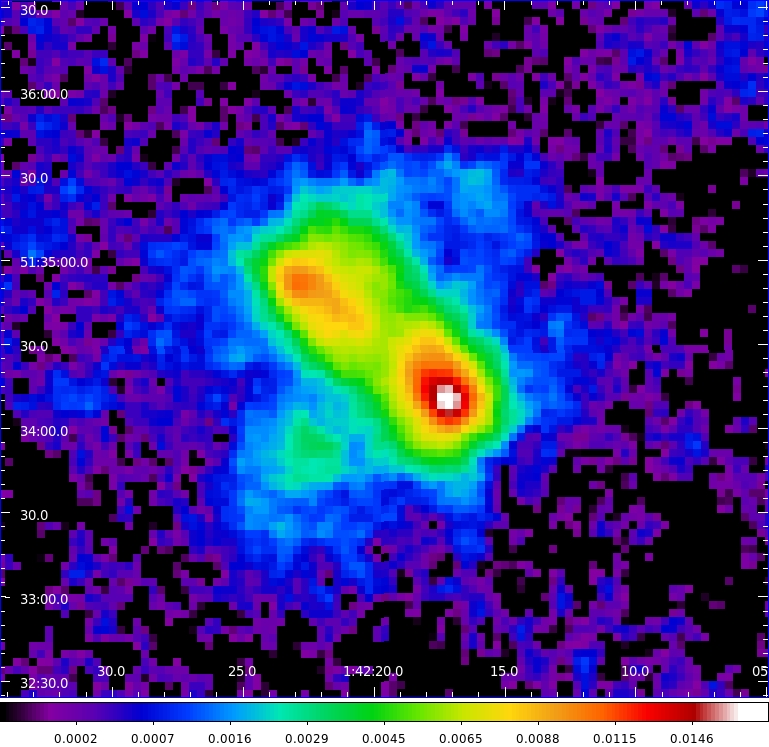} 
\caption{Same as Fig.~\ref{fig70} but showing the PACS 160~$\mu$m image.}
\label{fig160}
\end{center}
\end{figure}

\begin{figure}
\begin{center}
\includegraphics[width=\columnwidth]{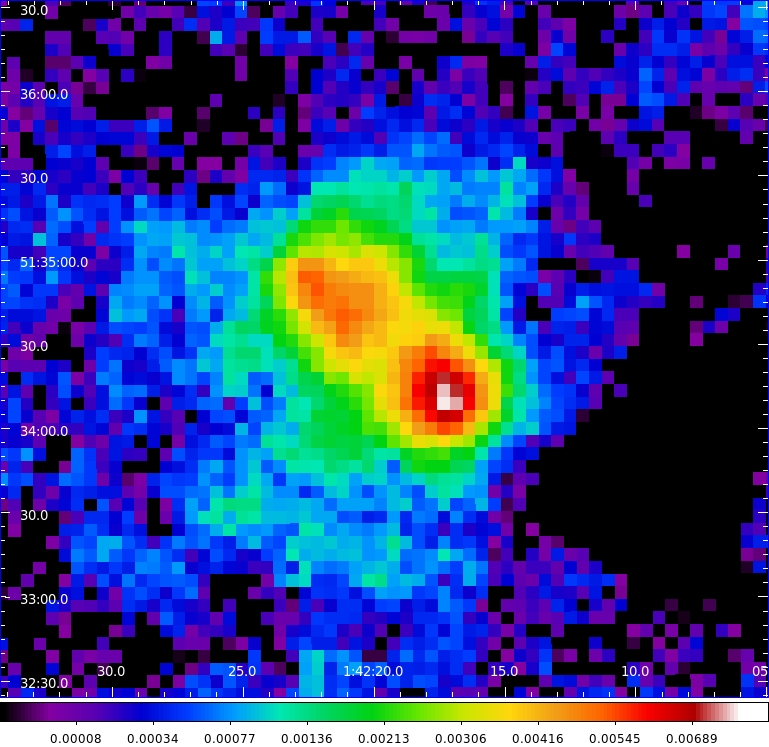} 
\caption{Same as Fig.~\ref{fig70} but showing the SPIRE 250~$\mu$m image.}
\label{fig250}
\end{center}
\end{figure}

\begin{figure}
\begin{center}
\includegraphics[width=\columnwidth]{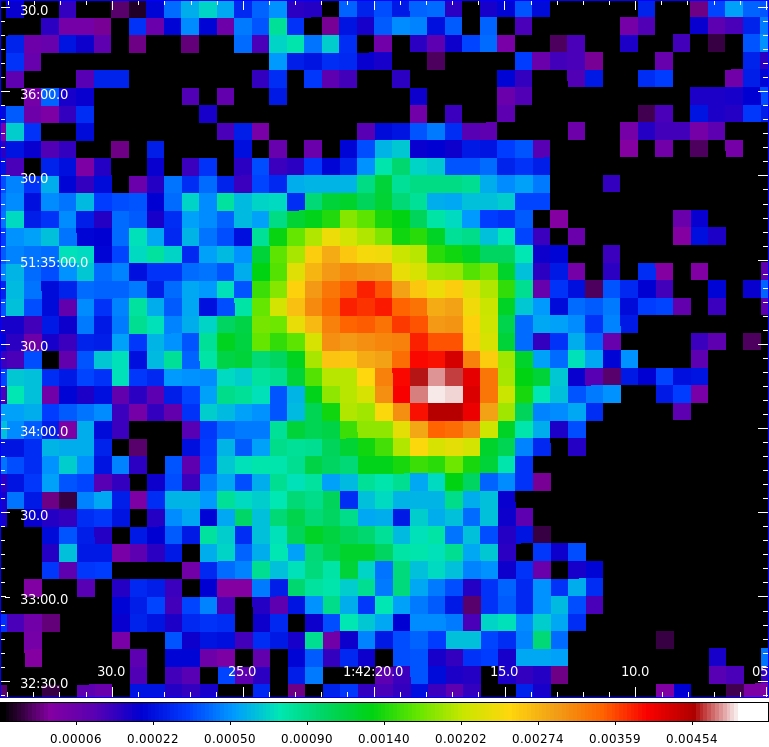} 
\caption{Same as Fig.~\ref{fig70} but showing the SPIRE 350~$\mu$m image.}
\label{fig350}
\end{center}
\end{figure}

\begin{figure}
\begin{center}
\includegraphics[width=\columnwidth]{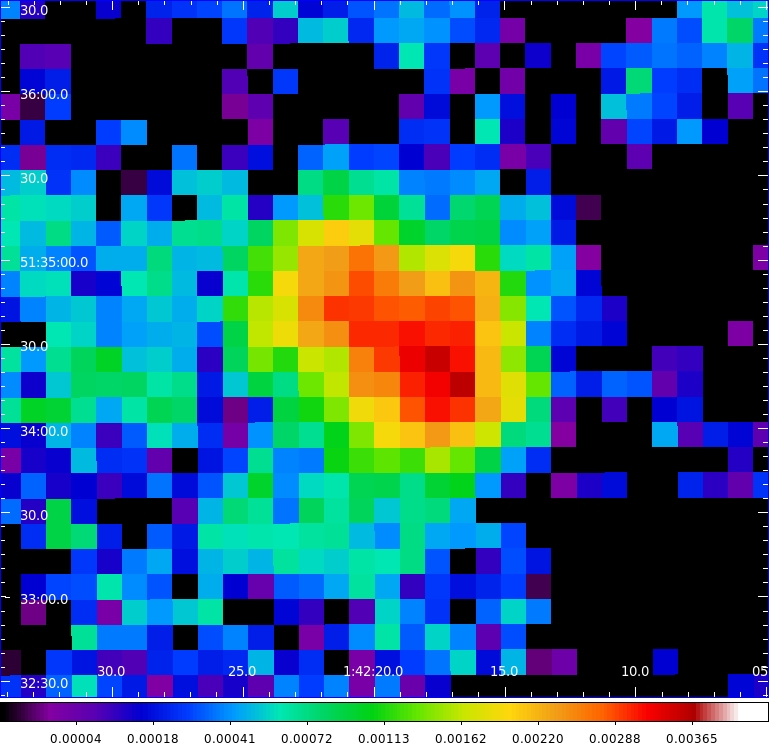} 
\caption{Same as Fig.~\ref{fig70} but showing the SPIRE 500~$\mu$m image.}
\label{fig500}
\end{center}
\end{figure}

NGC~650 has been imaged in scan map mode. With PACS we have obtained images in
the 70 and 160~$\mu$m bands, with SPIRE in the 250, 350, and 500~$\mu$m bands.
For PACS we used a scan speed of 20 arcsec/s and for SPIRE the nominal 30
arcsec/s. PACS and SPIRE data were reduced up to level~1 within the data
procession package HIPE \citep{Ott10} v9.0 and using the calibration files
PACS$\_$cal$\_$41 and SPIRE$\_$cal$\_$9.1, respectively. \ The PACS and SPIRE
images were made with the code Scanamorphos version 18 \citep{Roussel12}. We
used the standard pixel sizes in Scanamorphos which are 1.40~arcsec/pixel for
the 70~$\mu$m image, 2.85~arcsec/pixel for the 160 $\mu$m image, and 4.50,
6.25, 9.00 ~arcsec/pixel for SPIRE at 250, 350, 500 $\mu$m, respectively. The
Herschel beams have a complicated structure and are elongated as a result of
the scanning. The full width at half maximum (FWHM) is approximately
$6.0\times5.5$~arcsec for the 70~$\mu$m map and $12.0\times10.5$~arcsec for
the 160~$\mu$m map. The FWHM of the SPIRE beams are $18.3\times17.0$~arcsec
for 250~$\mu$m, $24.7\times23.2$~arcsec for 350~$\mu$m, and
$37.0\times33.4$~arcsec for 500~$\mu$m. The resulting images in Jy/pixel are
shown in Figs.~\ref{fig70}, \ref{fig160}, \ref{fig250}, \ref{fig350}, and
\ref{fig500}. The background sources were subtracted. Then the background was
measured over several regions around the object in the map. After background
subtraction, we measured fluxes within the signal-to-noise ratio = 1 contour
in a region around the object. The resulting fluxes are shown in the second
column of Table~\ref{irfluxes} combined with fluxes derived from IRAS (the
infrared astronomical satellite) (IPAC 1986) and Spitzer \citep{Ue06}
observations.

\begin{table}
\caption{The far infrared continuum fluxes measured by various satellite
  missions. The second column gives the quoted flux, the third column the
  conversion factors from point source calibration to extended source
  calibration for the SPIRE data, the fourth column the color correction
  factor, and the fifth column the actual flux with all corrections applied.
  See Sect~\ref{other:data} for details on the corrections.
\label{irfluxes}}
\begin{tabular}{lllll}
\hline
band & $F_\nu{\rm [quoted]}$ & $K_{\rm 4E}/K_{\rm 4P}$ & $K_{\rm color}$ & $F_\nu{\rm [actual]}$ \\
     & Jy                   &                      &               & Jy                    \\
\hline
IRAS 60       & 6.80$\pm$1.40 & 1.0000 & 0.8960 & 7.59$\pm$1.60 \\
PACS 70       & 6.95$\pm$0.70 & 1.0000 & 0.9749 & 7.13$\pm$0.70 \\
Spitzer 71.42 & 6.04$\pm$0.30 & 1.0000 & 0.8737 & 6.91$\pm$0.35 \\
IRAS 100      & 9.29$\pm$0.51 & 1.0000 & 0.9783 & 9.50$\pm$0.52 \\
Spitzer 155.9 & 4.83$\pm$0.94 & 1.0000 & 0.9863 & 4.90$\pm$0.95 \\
PACS 160      & 6.05$\pm$0.61 & 1.0000 & 1.0428 & 5.80$\pm$0.58 \\
SPIRE 250     & 1.50$\pm$0.16 & 0.9828 & 1.0173 & 1.45$\pm$0.15 \\
SPIRE 350     & 0.62$\pm$0.06 & 0.9834 & 1.0199 & 0.60$\pm$0.06 \\
SPIRE 500     & 0.24$\pm$0.04 & 0.9710 & 1.0094 & 0.23$\pm$0.04 \\
\hline
\end{tabular}
\end{table}


To produce flux ratio images, the background subtracted images were convolved
using the appropriate convolution kernels (May 2012) of \citet{Aniano11} and
rebinned to the pixel size of the longest wavelength image with flux
conservation in IDL using conv$\_$image.pro written by Karl D. Gordon at
STScI. To convert the flux ratio images to temperature maps, we determined the
theoretical flux ratio at a given grain temperature by folding the grain
emissivity of astronomical graphite \citep{Martin91} with the PACS and SPIRE
filter transmission curves in HIPE (v8.2.0) using the procedure outlined in
the SPIRE Observer's Manual. The use of graphite is justified by our finding
that the C/O ratio is larger than 1 (see Sect.~\ref{abundances}). We
interpolated the flux ratio as a function of temperature for each pixel in the
the flux ratio image to obtain the temperature map. This map is discussed in
Sect.~\ref{heating}. The procedure for making the temperature map is sensitive
to measurement uncertainties if both wavelengths are in the Rayleigh-Jeans
tail of the dust emission. Here we present only the temperature map based on
the PACS~70~/~160~$\mu$m ratio image. On the one hand, using the PACS
70~$\mu$m image assures that we get the best temperature determination since
this wavelength is shortward of the peak of the dust spectral energy
distribution (SED) implying that the flux is sensitive to the dust
temperature. On the other hand, using the PACS~160~$\mu$m image assures that
we get the best possible resolution in the temperature map since we need to
convolve the images to the resolution of the longest wavelength image prior to
taking the ratio. For PACS 160~$\mu$m the resulting FWHM is approximately
$12.0\times10.5$~arcsec, while for SPIRE 250~$\mu$m this would already have
been reduced to $18.3\times17.0$~arcsec. The images were not corrected for
line emission contributing to the inband flux as this cannot be done reliably.
The result is that the grain temperature will be slightly underestimated, by
about 1~K. See Sect.~\ref{sec:syn} for a further discussion of the line
contribution.

\begin{figure*}
\begin{center}
\includegraphics[width=\textwidth]{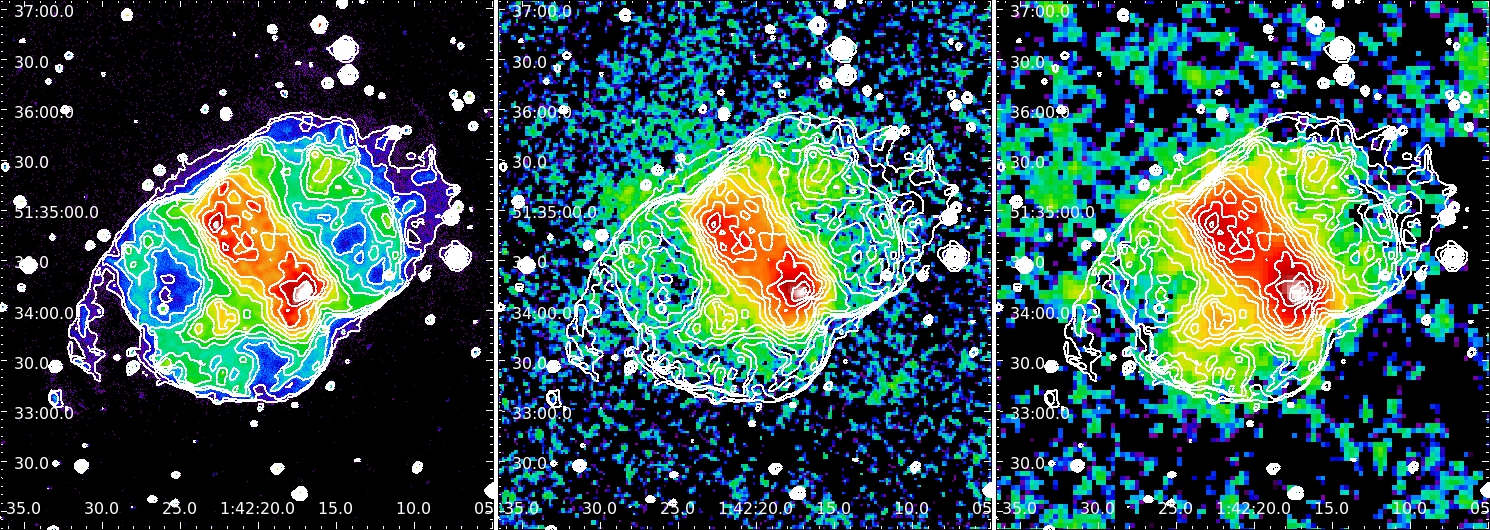} 
\caption{Left is the H$\alpha$ image from the IAC Catalog of Northern Galactic
  PNe \citep{Manchado96}. The contours of the H$\alpha$ image are overlaid on
  the PACS 70 and 160~$\mu$m images (middle and right panel, respectively).
  All images are displayed on a logarithmic scale to reveal the extent of the
  low level emission.
}
\label{figcomp}
\end{center}
\end{figure*}

In the PACS 70~$\mu$m data we can resolve the clumpy structure of the torus.
Because of the unprecedented high resolution of the PACS~70 $\mu$m image we
can compare it in Fig.~\ref{figcomp} to the Ha+[N II] band at
6600~\AA\ obtained from the ``The IAC morphological catalog of northern
galactic planetary nebulae'' \citep{Manchado96}. We see that the images agree
well: the brightest clumps of emission coincide in both images. Bright
H$\alpha$ emission and holes are visible at the edges in the northern lobe
towards the southwest end in the southern lobe towards the northeast. These
edges and holes are not discernable as such in the PACS images where there is
just faint dust emission across these regions. The fainter clumps we do not
see. In the PACS 160~$\mu$m image we confirm that there is dust emission in
these regions and all across the inner lobes. Likewise, the outer lobes and
even the fainter bow like structure at the the tips of the outer lobes are not
detected. We do not notice that the dust emission is more extended than the
H$\alpha$ emission as was suggested by \citet{Ue06}. This is most likely a
beam effect.

The infrared emission also correlates with the continuum subtracted H$_2$
image in \citet{ML13}. The H$_2$ emission comes from the knots and filaments
in the torus and some knots in the the two densest regions in the inner lobes.
Similarly, in NGC~6720 we also detect strong H$_2$ emission from regions with
strong dust emission, indicating that the dust and H$_2$ are correlated.

\begin{figure}
\begin{center}
\includegraphics[width=\columnwidth]{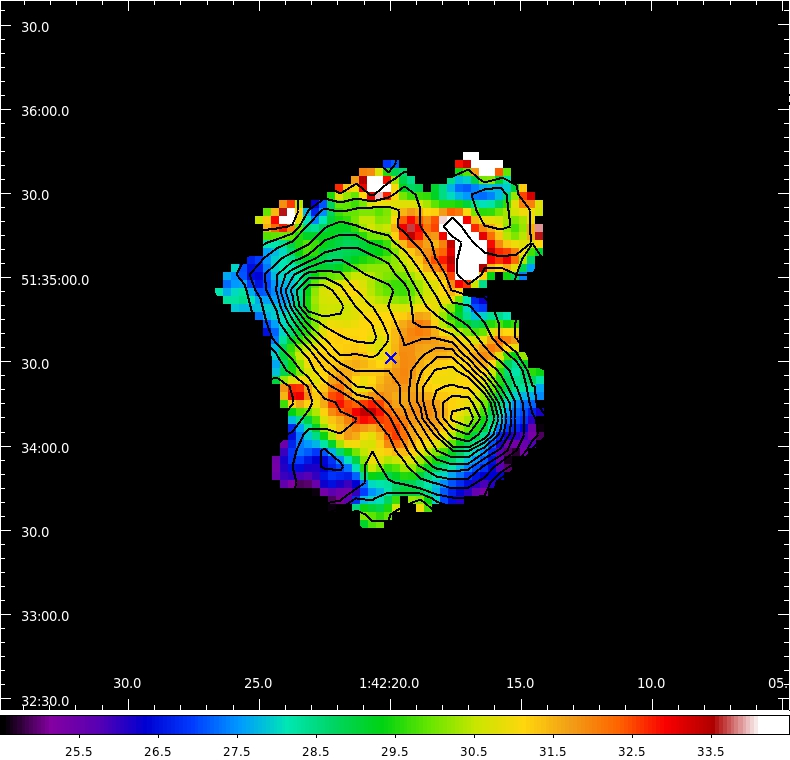} 
\caption{The temperature map of NGC 650 created from the PACS~70~/~160~$\mu$m
  ratio image. The black contours are taken from the PACS~160~$\mu$m image.
  The blue cross marks the location of the central star. The bar at the bottom
  shows the temperature scale.}
\label{figtemp}
\end{center}
\end{figure}

\begin{figure}
\begin{center}
\includegraphics[width=\columnwidth]{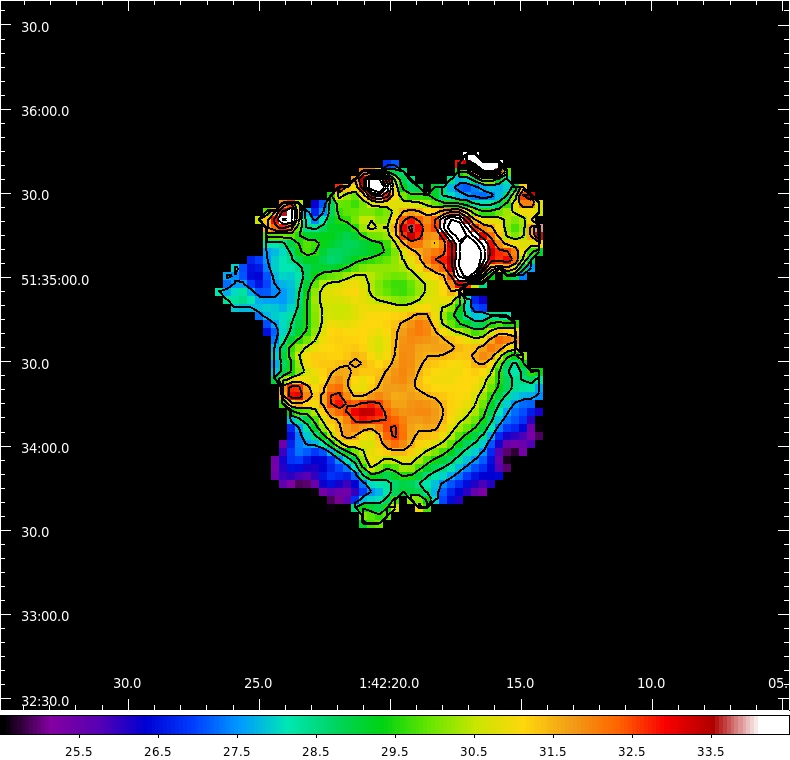} 
\caption{Same as Fig.~\ref{figtemp}, but showing contours for the grain
  temperature. Contour levels are from 27.5 to 35.5~K with 1~K increments.}
\label{figtempii}
\end{center}
\end{figure}

The temperature maps of NGC~650 are presented in Figs.~\ref{figtemp} and
\ref{figtempii}. In the temperature maps we do see that the highest
temperatures are observed at the position of the central star and towards the
brightest regions in the lobes. The temperature seems a bit higher to the
south. The temperature decreases towards the outer torus. We clearly see the
effect of the absorption of direct starlight on the temperature of the dust
due to which the dust is hotter in the low opacity regions close to the
central star and gets cooler as the extinction increases outwards through the
torus. This is discussed in Sect.~\ref{heating}.

\section{Cloudy modeling}
\label{model}

To interpret the imaging data we first created a photoionization model of
NGC~650. We modeled the nebula using the photoionization code Cloudy version
C10.01, last described by \citet{Fe98}. To determine the stellar and nebular
parameters, we used a method that was first described in \citet{vV99}. Below
are the main characteristics of the method.

The model for the PN is quite simple, and comprises the following assumptions:
\begin{enumerate}
\item{The spectral energy distribution of the central star is modeled using
  H-Ca PN model atmospheres from \citet{Ra97} for solar abundances.}
\item{The nebula has a toroidal geometry. In Cloudy this is approximated as a
  hollow sphere with the caps removed (this is the ``cylinder'' option in
  Cloudy).}
\item{The density is constant inside the ionized region.}
\item{Dust grains are intermixed with the gas at a constant dust-to-gas ratio.
  They are assumed to be single-sized and composed of graphite. The use of
  graphite is justified by our finding that the C/O ratio is larger than 1
  (see Sect.~\ref{abundances}).}
\item{The filling factor, describing the small scale clumpiness of the gas, is
  unity.}
\end{enumerate}
We chose to use the H-Ca model atmosphere grid rather than the newer PG 1159
or H-Ni grids \citep{Ra03} because the latter do not extend to high enough
temperatures for our needs.

The above assumptions leave the following free parameters: the stellar
temperature, the luminosity of the central star, the hydrogen density in the
ionized region, the inner radius of the nebula, the dust to gas mass ratio,
and the nebular abundances. The outer radius of the nebula is not fixed as an
input parameter. Instead the model is stopped when the observed flux density
at 350~$\mu$m is reached.

Adopting certain values for the input parameters, it is possible to calculate
a model for the nebula with Cloudy, predicting the continuum and line fluxes
as well as the Str\"omgren radius. To compare the model predictions with the
observed quantities, a non-standard goodness-of-fit estimator $\chi^2$ is
calculated. This estimator is minimized by varying all the input parameters of
the model, using the parallel algorithm Phymir which was specifically designed
for this task and is integrated in Cloudy.

It is assumed that there exists a unique set of input parameters for which the
resulting model predictions give the best fit to a given set of observables.
These input parameters are then considered the best estimate for the physical
properties of the PN.

The full set of observed quantities necessary to derive the physical
parameters of the PN are discussed below.
\begin{enumerate}
\item We need an emission line spectrum of the nebula, ranging from the UV to
  the IR. The line ratios make it possible to constrain the stellar
  temperature, the density and the electron temperature in the nebula. They
  are also required to determine the abundances. For elements for which no
  lines are available we assume standard abundances \citep{AC83, Kh89}.
\item Since dust is included in the model we also need information on the mid-
  and far-infrared continuum. For this various far-IR fluxes are used.
\item To constrain the emission measure, we use an optically thin radio
  continuum measurement at two different frequencies.
\item The angular diameter of the nebula is needed, which we define as
  $\Theta_{\rm d} = 2r_{\rm str}/D$. Here $r_{\rm str}$ stands for the
  Str\"omgren radius of the nebula and $D$ is the distance to the nebula.
\end{enumerate}
The specific data used to constrain the model of NGC 650 are described in more
detail in the following sections.

A more in-depth discussion of this method (including a description of how
$\chi^2$ is calculated) can be found in Chapter~2 of
\citet{vH97}\footnote{Available on-line at
  http://irs.ub.rug.nl/ppn/161821650.}

\subsection{The IUE data}
\label{sec:iue}

\begin{table*}
\caption{Log of the various IUE observations used in this paper. The position
  angle of the aperture is measured from north to east. The position number
  shown in column 6 corresponds to the numbering in Fig.~\ref{apertures}.
\label{iue}
}
\begin{tabular}{lrrrrrr}
\hline
ident & date & RA(J2000) & Dec(J2000) & P.A. & pos. & exp. time \\
      &      &    degree &     degree & degree &    &         s \\
\hline
swp32733 & 1988-01-16 & 25.59682 & 51.58220 & 328.03 & 2 &  5400 \\
swp38257 & 1990-02-25 & 25.57080 & 51.56806 & 301.84 & 6 & 14400 \\
swp42138 & 1991-07-29 & 25.59765 & 51.58660 & 141.51 & 1 & 27000 \\
swp42139 & 1991-07-29 & 25.57534 & 51.56945 & 141.26 & 5 &  8100 \\
swp42142 & 1991-07-30 & 25.58624 & 51.57803 & 140.70 & 3 & 12000 \\
swp42144 & 1991-07-31 & 25.58622 & 51.57303 & 140.26 & 4 & 14400 \\
lwp12498 & 1988-01-15 & 25.59682 & 51.58220 & 328.70 & 2 &  6300 \\
lwp17427 & 1990-02-26 & 25.57070 & 51.56806 & 301.08 & 6 &  7800 \\
lwp20907 & 1991-07-29 & 25.57534 & 51.56945 & 141.26 & 5 & 12000 \\
lwp20910 & 1991-07-30 & 25.59765 & 51.58660 & 140.86 & 1 & 21600 \\
lwp20912 & 1991-07-30 & 25.58624 & 51.57803 & 140.70 & 3 & 10200 \\
lwp20915 & 1991-07-31 & 25.58622 & 51.57303 & 140.26 & 4 & 12000 \\
\hline
\end{tabular}
\end{table*}

\begin{figure}
\includegraphics[width=1.00\columnwidth]{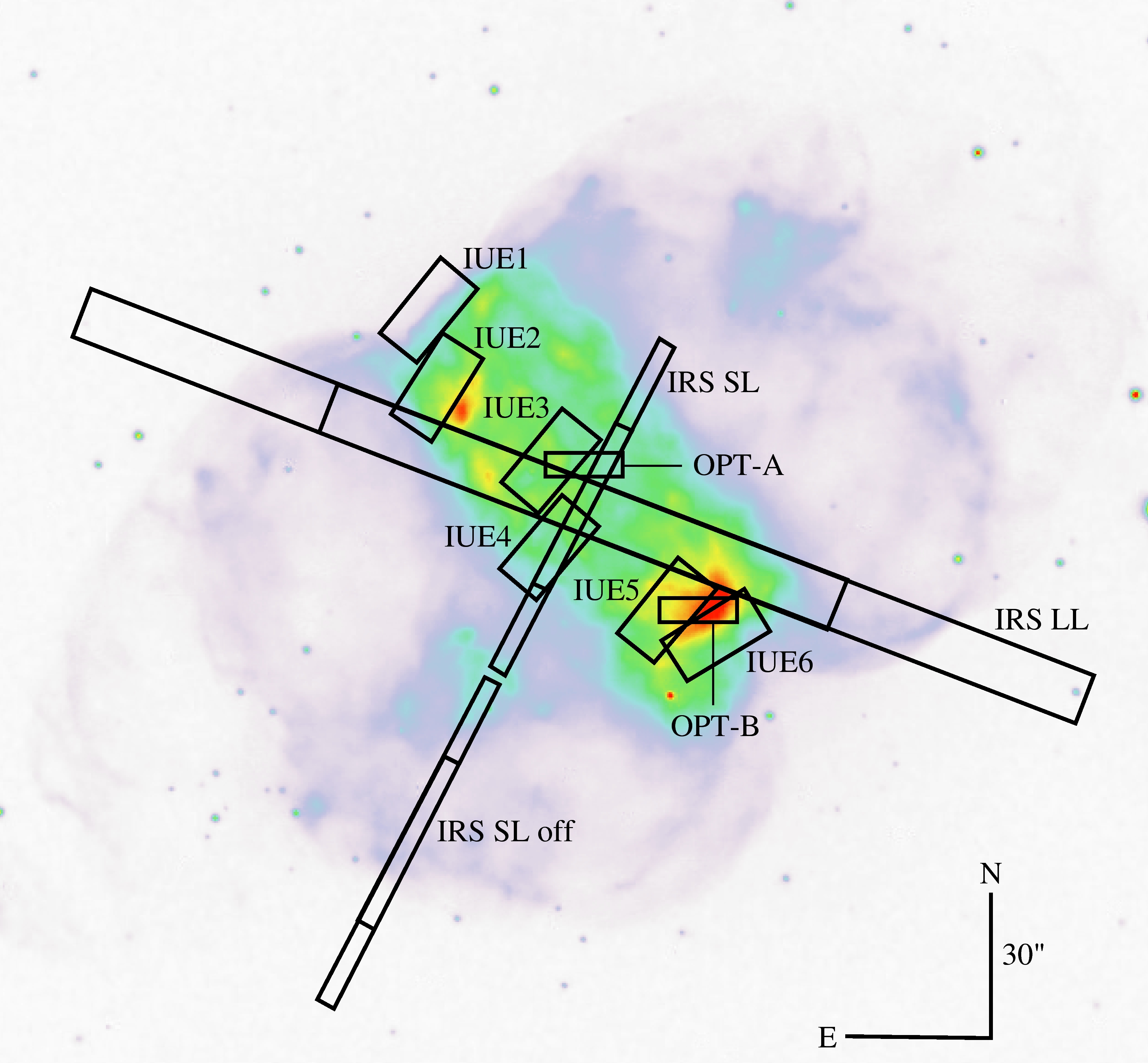}
\caption{The various apertures shown on top of the H$\alpha$ image taken with
  the NOT (credit: Lars \O. Andersen, Lars Malmgren, Frank R. Larsen).}
\label{apertures}
\end{figure}

\begin{table}
\caption{The line fluxes of NGC 650 observed with IUE. All wavelengths are
  given in vacuum. The fourth column gives the observed flux. The fifth column
  gives the flux relative to F(H$\beta$) = 100, the sixth column the
  dereddened relative flux, and the seventh column the relative flux predicted
  by the Cloudy model.
\label{iueflux}
}
{\footnotesize
\begin{tabular}{l@{\x}rl@{\x}r@{\x}r@{\x}r@{\x}r}
\hline
spectr. & $\lambda_{\rm lab}$ & mod. & F($\lambda$) & F$_{\rm n}$($\lambda$) & I($\lambda$) & I($\lambda$) \\
         & \AA              & \multicolumn{2}{r}{10$^{-17}$ W\,m$^{-2}$} &  &              & Cld          \\
\hline
C\,{\sc iv}     & 1549 & SWP &  31.1$\pm$2.6 &   41.$\pm$3.  & 215.$\pm$16.  &  267.  \\ \relax
He\,{\sc ii}    & 1640 & SWP &   58.$\pm$3.  &   77.$\pm$4.  & 364.$\pm$19.  &  419.  \\ \relax
O\,{\sc iii}]   & 1665 & SWP &   6.6$\pm$1.6 &   8.8$\pm$2.1 &  41.$\pm$10.  &   21.  \\ \relax
N\,{\sc iii}]   & 1750 & SWP &   7.1$\pm$1.5 &   9.4$\pm$2.0 &  42.$\pm$9.   &   31.  \\ \relax
C\,{\sc iii}]   & 1909 & SWP & 131.8$\pm$2.5 &  175.$\pm$3.  & 802.$\pm$14.  & 1023.  \\ \relax
C\,{\sc ii}]    & 2326 & LWP &  56.8$\pm$2.6 &   75.$\pm$3.  & 339.$\pm$14.  &  275.  \\ \relax
[Ne\,{\sc iv}]  & 2424 & LWP &   9.0$\pm$1.3 &  11.9$\pm$1.7 &  41.$\pm$6.   &   34.  \\ \relax
[O\,{\sc ii}]   & 2471 & LWP &   6.6$\pm$1.5 &   8.8$\pm$2.0 &  28.$\pm$6.   &   12.  \\ \relax
He\,{\sc ii}    & 2734 & LWP &   5.1$\pm$0.8 &   6.8$\pm$1.1 &  15.$\pm$3.   &   13.  \\ \relax
Mg\,{\sc ii}    & 2796 & LWP &   2.8$\pm$0.5 &   3.7$\pm$0.7 &  7.8$\pm$1.5  &    7.4 \\ \relax
[Ar\,{\sc iii}]?&3110\m& LWP &   5.0$\pm$1.5 &   6.6$\pm$2.0 &  11.$\pm$3.   &    0.9 \\ \relax
He\,{\sc ii}    & 3204 & LWP &   6.4$\pm$1.7 &   8.5$\pm$2.3 &  14.$\pm$4.   &   25.  \\
\hline
\multicolumn{7}{l}{\m\ \ This line was not used to constrain the model.}
\end{tabular}
}
\end{table}

\begin{figure}
\includegraphics[width=1.00\columnwidth]{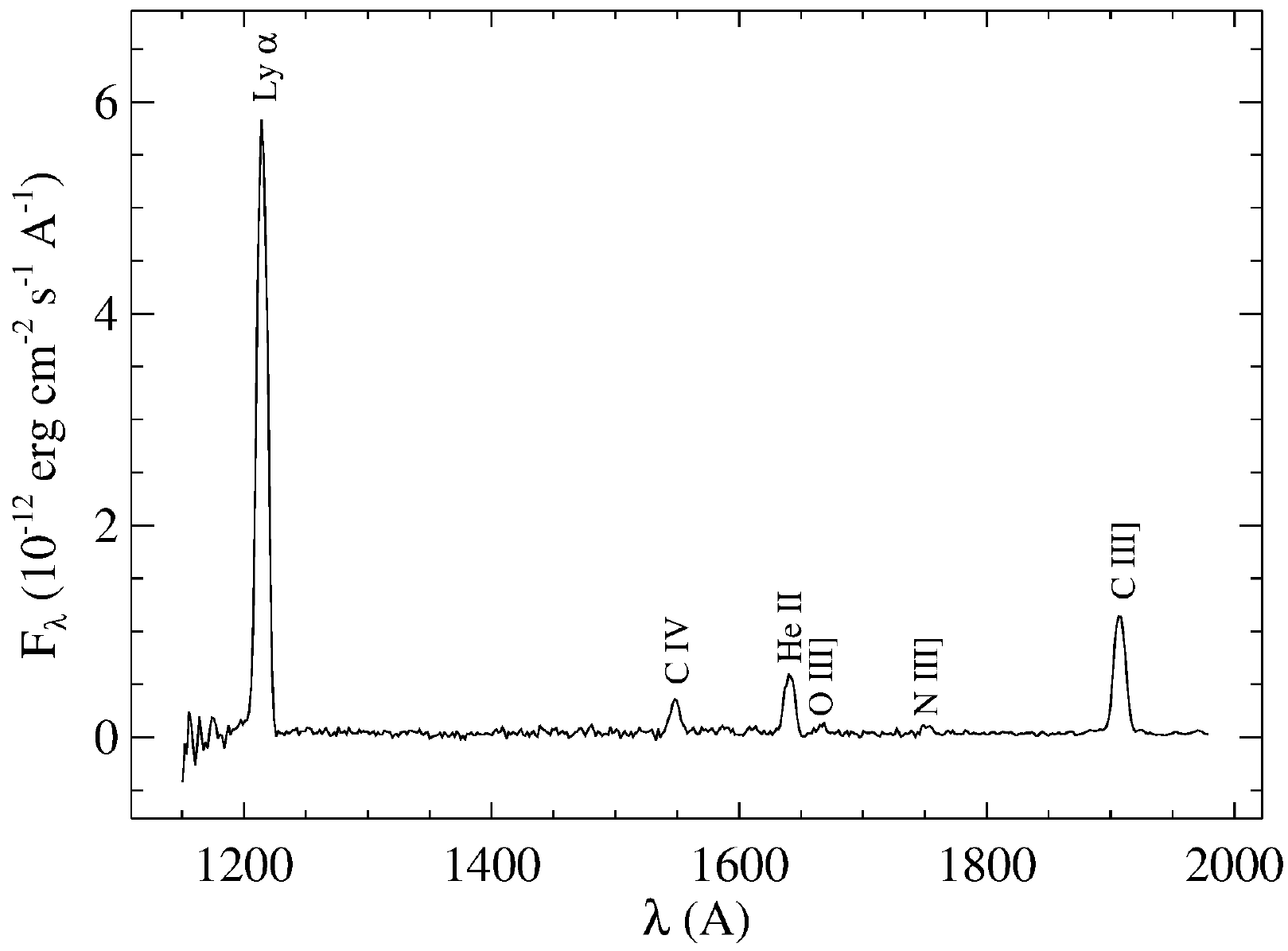}
\includegraphics[width=1.00\columnwidth]{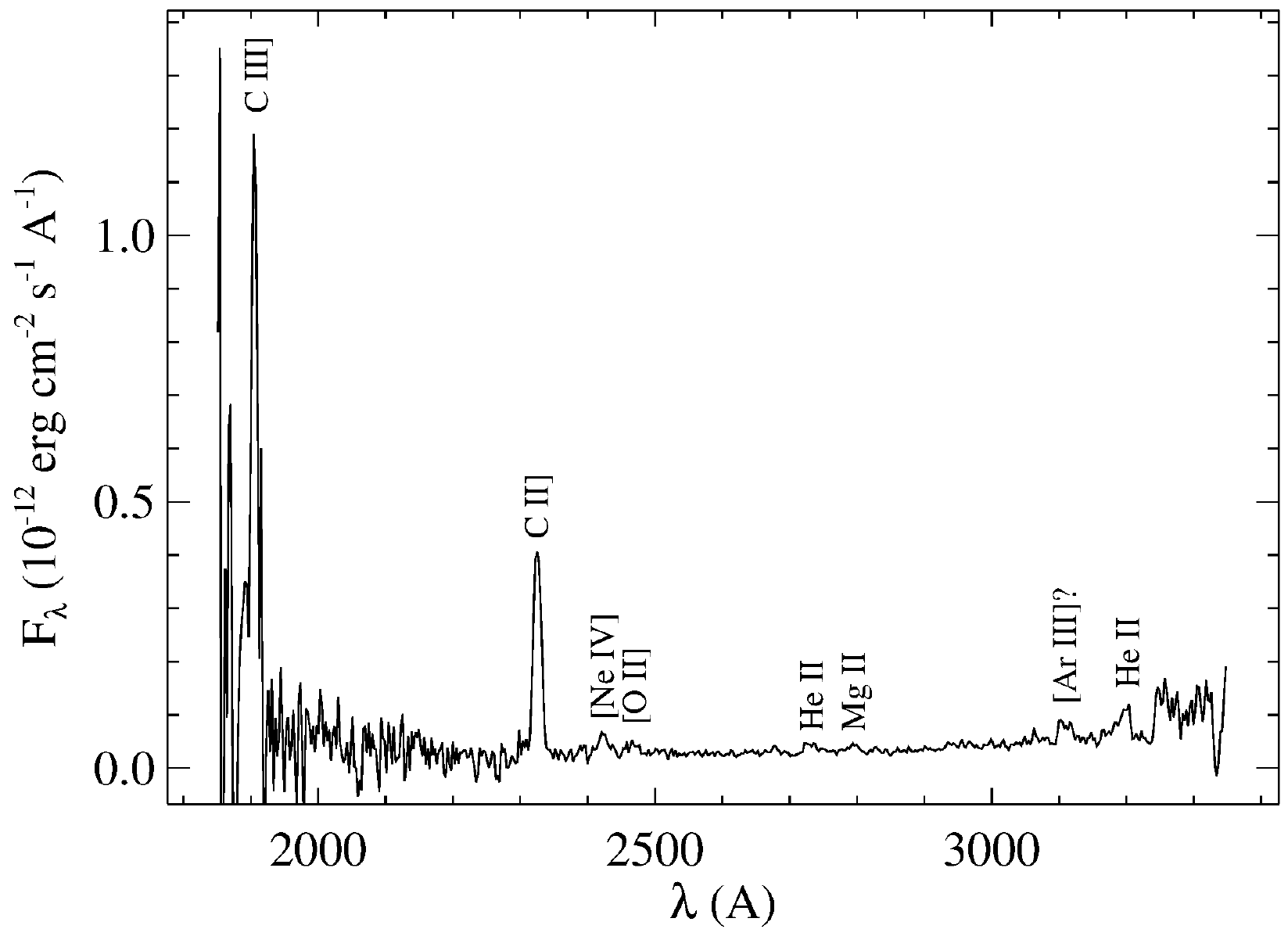}
\caption{The IUE spectra of NGC 650. The top panel shows the SWP spectrum,
  while the bottom panel shows the LWP spectrum.}
\label{iuefig}
\end{figure}

Many large-aperture IUE (international ultraviolet explorer) 
spectra of NGC~650 exist, covering the short and long
wavelength section. The spectra sample various positions along the central bar
as well as one of the lobes of the PN. We decided to use all spectra that were
pointed at the bar and calculate a weighted average of these to approximate a
long-slit spectrum over the central bar. The observations that we used are
summarized in Table~\ref{iue}, the aperture positions are shown in
Fig.~\ref{apertures}. We used the standard pipeline reductions from the MAST
IUE database and analyzed these using the Starlink SPLAT-VO tool version
3.9-6. First we averaged the spectra giving the spectra at positions 2 through
5 a weight of 2 and the spectra at positions 1 and 6 a weight of 1. This was
done to approximately compensate for the missing flux between positions 2 and
3, as well as 4 and 5. The resulting spectra are shown in Fig.~\ref{iuefig}.
We measured the line fluxes in the averaged spectrum. These are shown in
Table~\ref{iueflux}. All lines except the 4 brightest ones should be
considered uncertain. These are only marginal detections and the uncertainties
returned by SPLAT-VO are likely too optimistic. We could not confirm the
presence of the C\,{\sc ii} 1336~\AA\ and O\,{\sc iv}] 1404~\AA\ lines
  reported by \citet{KH96}.

\begin{table}
\caption{The line fluxes of NGC 650 resulting from combining the two spectra
  from \citet{Kw03} as discussed in Sect.~\ref{opt:spectrum}. All wavelengths
  are given in air. The third and fourth columns give the flux relative to
  F(H$\beta$) = 100 for each of the apertures, the fifth column the dereddened
  relative flux of the combined spectrum, and the sixth column the relative
  flux predicted by the Cloudy model. Line fluxes marked with a colon are
  uncertain.
\label{optflux}
}
\begin{tabular}{lrrrrr}
\hline
spectrum & $\lambda_{\rm lab}$ & F$_{\rm n}$($\lambda$) & F$_{\rm n}$($\lambda$) & I($\lambda$) & I($\lambda$) \\
         & \AA               & OPT-A                 & OPT-B                 &              & Cld          \\
\hline
[O\,{\sc ii}]                 & 3727 & 237.  & 453.  &  525.  & 699.  \\ \relax
H\,{\sc i}                    & 3798 & 3.0\uu& 3.3   &    3.8 &   5.2 \\ \relax
H\,{\sc i}                    & 3835 & 8.0   & 7.6   &    8.7 &   7.2 \\ \relax
[Ne\,{\sc iii}]               & 3869 & 126.  & 144.  &  164.  & 166.  \\ \relax
H\,{\sc i} + He\,{\sc i}      &3889\m& 17.4  & 18.0  &   20.4 &  17.7 \\ \relax
[Ne\,{\sc iii}] + H\,{\sc i}  &3968\m& 85.9  & 94.4  &  105.9 &  65.8 \\ \relax
He\,{\sc i} + He\,{\sc ii}    & 4026 & 1.1\uu& 1.6   &    1.8 &   1.5 \\ \relax
[S\,{\sc ii}]                 &4072\m& 4.4   & 7.1   &    7.9 &   9.5 \\ \relax
H\,{\sc i}                    & 4102 & 21.0  & 20.6  &   22.7 &  25.9 \\ \relax
He\,{\sc ii}                  & 4200 & 0.9\uu& 0.3\uu& 0.9\uu &   1.0 \\ \relax
C\,{\sc ii}                   &4267\m& 0.6   & 0.6\uu&    0.6 &   0.6 \\ \relax
H\,{\sc i}                    & 4340 & 43.4  & 42.9  &   45.9 &  46.9 \\ \relax
[O\,{\sc iii}]                & 4363 & 10.0  & 8.8   &   10.2 &  13.9 \\ \relax
He\,{\sc i}                   & 4471 & 2.2   & 3.3   &    3.5 &   3.1 \\ \relax
He\,{\sc ii}                  & 4542 & 1.6   & 0.9\u &    1.6 &   2.0 \\ \relax
He\,{\sc ii}                  & 4686 & 54.6  & 27.9  &   55.1 &  65.1 \\ \relax
[Ar\,{\sc iv}]                &4711\m& 3.4   & 1.2   &    3.4 &   3.5 \\ \relax
[Ar\,{\sc iv}]                & 4740 & 2.0   & 0.3   &    2.0 &   2.6 \\ \relax
H\,{\sc i}                    & 4861 & 100.  & 100.  &  100.  & 100.  \\ \relax
He\,{\sc i}                   & 4922 & 0.5   & 1.0   &    1.0 &   0.8 \\ \relax
[O\,{\sc iii}]                & 4959 & 339.  & 293.  &  337.  & 356.  \\ \relax
[O\,{\sc iii}]                & 5007 & 1123. & 961.  & 1115.  &1071.  \\ \relax
[N\,{\sc i}]                  & 5199 & 3.2   & 10.0  &    9.6 &  14.3 \\ \relax
He\,{\sc ii}                  & 5412 & 4.9   & 2.2   &    4.8 &   4.6 \\ \relax
[Cl\,{\sc iii}]               & 5518 & 0.9   & 1.0   &    0.9 &   0.7 \\ \relax
[O\,{\sc i}]                  &5577\m& ---   & 0.5   &    0.5 &   0.7 \\ \relax
[N\,{\sc ii}]                 & 5755 & 5.5   & 11.2  &   10.1 &  10.0 \\ \relax
He\,{\sc i}                   & 5876 & 7.5   & 12.0  &   10.7 &   8.3 \\ \relax
[O\,{\sc i}]                  & 6300 & 14.6  & 39.9  &   34.3 &  32.4 \\ \relax
[S\,{\sc iii}]                & 6312 & 3.9\u & 3.6   &  3.7\u &   3.8 \\ \relax
[O\,{\sc i}]                  & 6364 & 4.4\u & 11.9  &   10.2 &  10.3 \\ \relax
[N\,{\sc ii}]                 & 6548 & 98.1  & 219.  &  185.0 & 178.  \\ \relax
H\,{\sc i}                    & 6563 & 306.  & 340.  &  287.  & 290.  \\ \relax
[N\,{\sc ii}]                 & 6583 & 310.  & 672.  &  566.  & 527.  \\ \relax
He\,{\sc i}                   & 6678 & 2.4   & 3.7   &    3.1 &   2.3 \\ \relax
[S\,{\sc ii}]                 & 6716 & 37.8  & 71.5  &   59.8 &  56.7 \\ \relax
[S\,{\sc ii}]                 & 6731 & 30.1  & 59.2  &   49.5 &  45.3 \\ \relax
He\,{\sc ii}                  &6891\m& 0.5\uu& 0.4\uu& 0.5\uu &   0.5 \\ \relax
[Ar\,{\sc v}]                 & 7005 & 0.8   & ---   &    0.7 &   0.6 \\ \relax
He\,{\sc i}                   & 7065 & 1.8   & 3.2   &    2.6 &   1.7 \\ \relax
[Ar\,{\sc iii}]               & 7136 & 29.6  & 38.5  &   31.5 &  31.1 \\ \relax
He\,{\sc ii}                  & 7178 & 0.7   & 0.4\u &    0.6 &   0.7 \\ \relax
[Ar\,{\sc iv}]?               &7237\m& 0.6   & 0.4\u &    0.6 &  0.06 \\ \relax
He\,{\sc i}                   & 7281 & 0.5   & 0.5   &    0.5 &   0.4 \\ \relax
[O\,{\sc ii}]                 & 7323 & 8.4   & 19.7  &   16.0 &  15.8 \\ \relax
He\,{\sc ii}                  & 7593 & 1.0   & 0.7\uu&    0.9 &   0.9 \\ \relax
[Ar\,{\sc iii}]               & 7751 & 7.3   & 9.1   &    7.3 &   7.5 \\ \relax
[Cl\,{\sc iv}]                & 8046 & 0.6\u & 0.2\u &  0.5\u &   0.7 \\ \relax
He\,{\sc ii}                  & 8237 & 1.6   & 1.1   &    1.5 &   1.3 \\ \relax
H\,{\sc i}                    & 8665 & 1.3\u & 1.6\u &  1.2\u &   0.8 \\ \relax
H\,{\sc i}                    & 8750 & 1.2\u & 1.6\u &  1.2\u &   1.0 \\ \relax
H\,{\sc i}                    & 8863 & 1.9\u & 2.0\u &  1.7\u &   1.3 \\ \relax
H\,{\sc i}                    & 9015 & 2.0\u & 2.7\u &  2.1\u &   1.8 \\ \relax
[S\,{\sc iii}]                &9069\m& 44.7  & 52.7  &   40.5 &  35.6 \\ \relax
H\,{\sc i}                    & 9229 & 4.9\u & 4.7\u &  4.4\u &   2.5 \\ \relax
[S\,{\sc iii}]                & 9531 & 109.  & 187.  &  142.  &  88.4 \\ \relax
H\,{\sc i}                    & 9546 & 5.0\u & 7.9\u &  6.0\u &   3.5 \\
\hline
\multicolumn{6}{l}{\m\ \ This line was not used to constrain the model.}
\end{tabular}
\vspace*{-1em}
\end{table}

To calculate the ratio of the line fluxes to F(H$\beta$), we calculated the
average H$\beta$ flux covered by each of the IUE apertures using the H$\beta$
image of NGC~650 \citep{RL08} and using the same weighting discussed above.
After astrometrically calibrating the image and subtracting the background
emission as well as the field stars, we determined that the average IUE slit
covers a fraction 0.0361 of the total H$\beta$ flux. We created a custom-built
IDL routine for this purpose. The total observed H$\beta$ flux of the PN is
log F(H$\beta$) = -13.68~W\,m$^{-2}$ \citep{Ka78}. So the H$\beta$ flux
``covered'' by the average IUE aperture is 7.54$\times 10^{-16}$~W\,m$^{-2}$.
Note that this flux is a factor two higher than the one assumed by
\citet{KH96}. They used the IUE6 aperture, which partially covers the region
with the highest H$\beta$ surface brightness. Using the H$\beta$ image we
checked that the H$\beta$ flux covered by the IUE6 aperture alone is very
close to the average H$\beta$ flux quoted above (within 2\%). \citet{KH96}
used the H$\beta$ flux from an optical spectrum observed by \citet{PT87} which
was taken very close to the IUE6 aperture. However, apparently they did not
correct for the fact that the optical aperture was much smaller than the IUE
aperture (47~arcsec$^2$ vs. 200~arcsec$^2$). This choice implies that the IUE
line ratios and hence also the abundances based on the IUE lines reported by
\citet{KH96} are too large. In particular this is the case for the carbon
abundance and the C/O ratio (the oxygen abundance is largely based on optical
lines). The dereddening is discussed in Sect.~\ref{dered}.


\subsection{The optical data}
\label{opt:spectrum}

Obtaining optical spectra of NGC 650 that we can use in our modeling is quite
difficult. We are not aware of any spectra covering (a sizeable fraction of)
the bar, or even a long-slit spectrum along the bar. We could only find
spectra covering small areas of the nebula, almost like a pencil beam. This is
not ideal given that there is pronounced ionization stratification in the
nebula and the spectra will not measure what we predict in our Cloudy modeling
(which is the integrated flux from the entire nebula). We decided to use the
optical spectra from \citet{Kw03}. They obtained two spectra which we call
OPT-A and OPT-B. The position of the slits are shown in Fig.~\ref{apertures}.
\citet{Kw03} used only the central 16\arcsec\ from the slit, so only that part
of the slit is shown. OPT-A was taken near the central star and is clearly
biased towards higher ionization stages in the nebula. OPT-B was taken on the
bright area in the SW of the bar. This spectrum is clearly biased towards
lower ionization stages in the nebula. So neither spectrum can be compared to
the Cloudy model as such. We attempted using the sum of both apertures, but
this procedure also gave unsatisfactory results as the flux in the OPT-B
aperture is much higher and the summed spectrum is still clearly biased
towards lower ionization stages. Using preliminary Cloudy runs we noticed that
the predicted flux ratio for the He\,{\sc ii} 4686~\AA\ line was nearly
constant with a value very close to the observed ratio in OPT-A. This is
because this ratio is set by the central star temperature (more precisely the
number of He$^{2+}$ ionizing photons emitted by the central star) which is
well constrained by the observed spectrum. A close match between the observed
and modeled He\,{\sc ii} 4686~\AA\ line is very important since this line
couples the optical and IUE spectra. This is discussed in more detail in
Sect.~\ref{dered}. On the other hand we noted that low excitation lines like
e.g.\ [O\,{\sc ii}] 3727~\AA\ were consistently predicted very strong, with
model fluxes close to, or often even exceeding the observed ratio in the OPT-B
spectrum. Based on these observations we adopted the following ad hoc
procedure. We dereddened each of the spectra using the procedure described in
Sect.~\ref{dered}. For each line we then chose the maximum value for the
dereddened relative flux from each of the spectra. This procedure avoids a
bias towards either high- or low-excitation areas in the nebula. The resulting
line fluxes are listed in Table~\ref{optflux}.

\subsection{The Spitzer data}

\begin{figure}
\includegraphics[width=1.00\columnwidth]{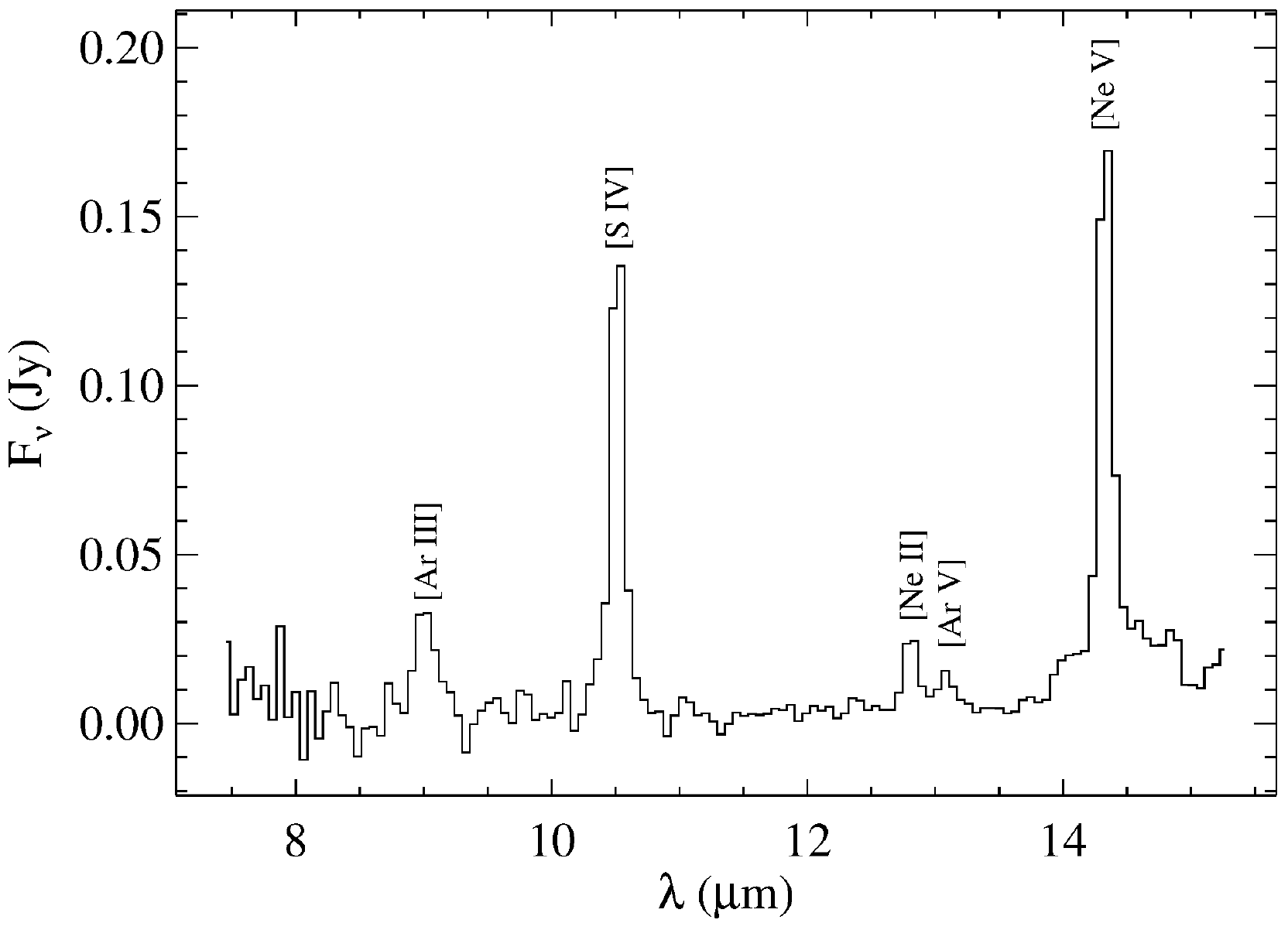}
\includegraphics[width=1.00\columnwidth]{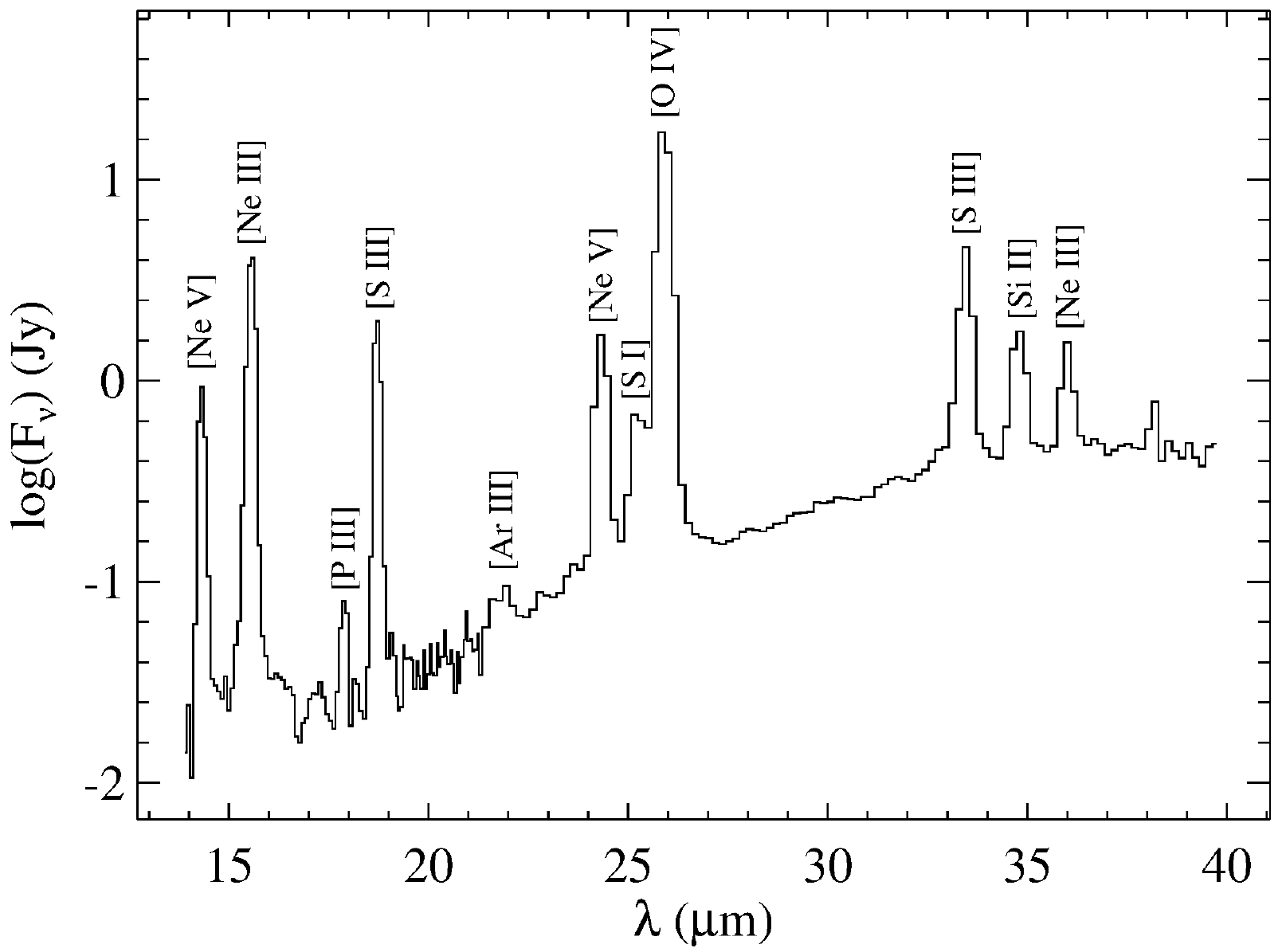}
\caption{The Spitzer IRS spectra of NGC 650. The top panel shows the SL1
  spectrum, while the bottom panel shows the combined LL1 and LL2 spectra. The
  broad feature underneath the [Ne\,{\sc v}] 14.3~$\mu$m line in the SL1
  spectrum is an artifact of the detector -- the so-called ``teardrop''. The
  LL spectra are shown on a logarithmic scale to emphasize the continuum
  emission.}
\label{irsfig}
\end{figure}

For the far-IR fine-structure lines we used the unpublished low-resolution
Spitzer IRS spectra 25691392 obtained in staring mode on 2009-03-08. The
Spitzer Space Telescope is described in \citet{We04}. The IRS instrument is
described in \citet{Ho04}. The spectra were reduced using the SMART package
v8.2.5 \citep{Hi04,Le10}. We used standard reduction steps for the data
preparation: the individual files were combined into 3-plane data, the data
were cleaned using IRSCLEAN with standard settings, and finally the data for
each nod were co-added. For the low resolution spectrum we subtracted the SL1
from the SL2 data (and vice versa) for each nod to do the low-level rogue
pixel removal (subtraction by order). We used the same procedure for the LL1
and LL2 data. Subtraction by nod could not be used since the slit is mostly or
completely filled by the source. Such a procedure would lead to significant
cancellation of the line fluxes. In the case of the SL spectra using
subtraction by order will also lead to a slight cancellation effect since the
``off-target'' slit will also pick up some flux from the faint lobes (see
Fig.~\ref{apertures}). We estimate that the resulting error in the flux should
not exceed 10 -- 20\% of the reported flux. For the LL spectra this is not an
issue since there the off-target slit is indeed completely off target. We
extracted the spectra using the ``Full aperture (extended source)'' mode.
Finally we merged the two nods by calculating the mean. The resulting spectra
are shown in Fig.~\ref{irsfig}. The emission lines were identified using the
Atomic Line List\footnote{http://www.pa.uky.edu/\textasciitilde peter/atomic}.
The blended [S\,{\sc i}] and [O\,{\sc iv}] lines were measured using
multi-component gaussian fits in SMART. The measured line fluxes are given in
Table~\ref{irs}.

To calculate the ratio of the line fluxes to F(H$\beta$), we calculated the
H$\beta$ flux covered by the SL and LL apertures using the same procedure as
outlined in Sect.~\ref{sec:iue}. The SL slit covers a fraction 0.0242 of the
total H$\beta$ flux averaged over the two nod positions. For the LL slit this
fraction is 0.186. This implies that the H$\beta$ flux ``covered'' by the IRS
slits is 5.06$\times 10^{-16}$~W\,m$^{-2}$ for the average SL slit and
3.89$\times 10^{-15}$~W\,m$^{-2}$ for the average LL slit.

\begin{table}
\caption{The line fluxes of NGC 650 observed with Spitzer IRS. All wavelengths
  are given in vacuum. The fourth column gives the observed flux. The fifth
  column gives the flux relative to F(H$\beta$) = 100, the sixth column the
  dereddened relative flux, and the seventh column the relative flux predicted
  by the Cloudy model.
\label{irs}
}
{\footnotesize
\begin{tabular}{l@{\x}rl@{\x}r@{\x}r@{\x}r@{\x}r}
\hline
spectr. & $\lambda_{\rm lab}$ & mod. & F($\lambda$) & F$_{\rm n}$($\lambda$) & I($\lambda$) & I($\lambda$) \\
        & $\mu$m & \multicolumn{2}{r}{10$^{-17}$ W\,m$^{-2}$} &             &              & Cld          \\
\hline
[Ar\,{\sc iii}] &  8.991 & SL1 &  24.$\pm$4.   &  47.$\pm$8.   &  22.$\pm$4.   &  20.  \\ \relax
[S\,{\sc iv}]   & 10.510 & SL1 &  54.$\pm$4.   & 107.$\pm$8.   &  51.$\pm$4.   &  55.  \\ \relax
[Ne\,{\sc ii}]  & 12.814 & SL1 &  5.7$\pm$0.5  & 11.3$\pm$1.0  &  5.2$\pm$0.5  &   4.9 \\ \relax
[Ar\,{\sc v}]   & 13.102 & SL1 &  3.0$\pm$0.5  &  5.9$\pm$1.0  &  2.7$\pm$0.5  &   2.3 \\ \relax
[Ne\,{\sc v}]   & 14.322 & SL1 & 30.5$\pm$0.8  & 60.3$\pm$1.6  & 27.6$\pm$0.7  &  27.0 \\ \relax
[Ne\,{\sc v}]   & 14.322 & LL2 & 267.$\pm$15.  &  69.$\pm$4.   & 31.6$\pm$1.8  &  27.0 \\ \relax
[Ne\,{\sc iii}] & 15.555 & LL2 &1160.$\pm$30.  & 298.$\pm$8.   & 136.$\pm$4.   & 130.  \\ \relax
[P\,{\sc iii}]  & 17.885 & LL2 & 10.5$\pm$2.3  &  2.7$\pm$0.6  &  1.2$\pm$0.3  &   1.1 \\ \relax
[S\,{\sc iii}]  & 18.713 & LL2 & 342.$\pm$26.  &  88.$\pm$7.   &  41.$\pm$3.   &  37.  \\ \relax
[Ar\,{\sc iii}] &21.829\m& LL1 &  7.7$\pm$1.3  &  2.0$\pm$0.3  & 0.91$\pm$0.15 &  1.44 \\ \relax
[Ne\,{\sc v}]   & 24.318 & LL1 & 274.$\pm$11.  & 70.4$\pm$2.8  & 32.1$\pm$1.3  &  33.0 \\ \relax
[S\,{\sc i}]    &25.249\m& LL1 & 110.$\pm$30.  &  28.$\pm$8.   &  13.$\pm$4.   &   0.03\\ \relax
[O\,{\sc iv}]   & 25.890 & LL1 &2850.$\pm$60.  & 733.$\pm$15.  & 334.$\pm$7.   & 323.  \\ \relax
[S\,{\sc iii}]  & 33.481 & LL1 & 371.$\pm$14.  &  95.$\pm$4 .  & 43.0$\pm$1.8  &  67.9 \\ \relax
[Si\,{\sc ii}]  & 34.815 & LL1 & 141.$\pm$5.   & 36.2$\pm$1.3  & 16.4$\pm$0.6  &  17.4 \\ \relax
[Ne\,{\sc iii}] & 36.014 & LL1 &  81.$\pm$3.   & 20.8$\pm$0.8  &  9.4$\pm$0.4  &  11.7 \\
\hline
\multicolumn{7}{l}{\m\ \ This line was not used to constrain the model.}
\end{tabular}
}
\end{table}

\subsection{Other data}
\label{other:data}

\begin{figure}
\includegraphics[width=1.00\columnwidth]{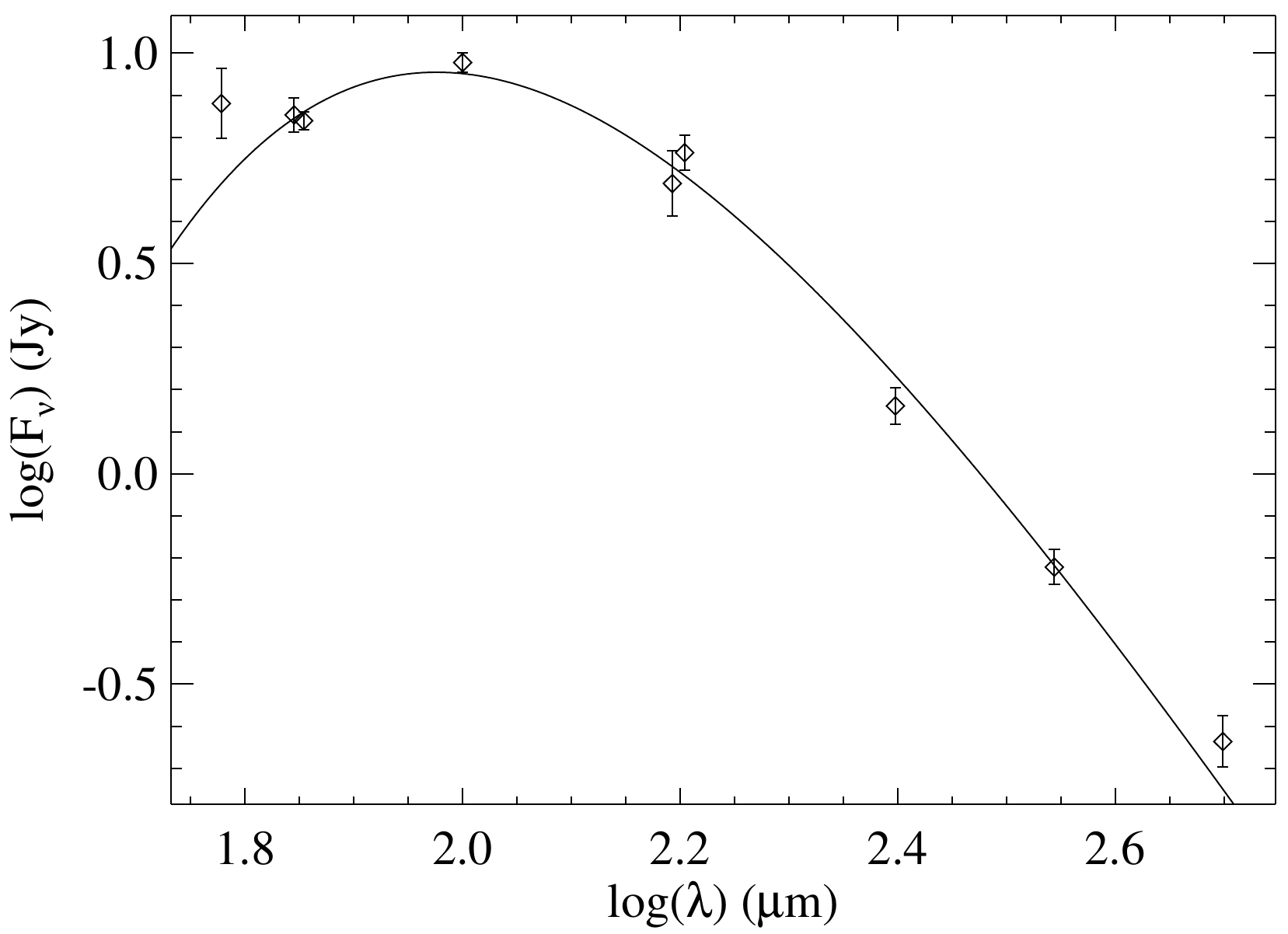}
\caption{The SED of the dust in NGC 650. The solid line shows the modified
  blackbody that was fitted to the observations.}
\label{sed}
\end{figure}

To constrain the free-free emission from the ionized gas, we used a 4.85~GHz
measurement of 114$\pm$10~mJy from the GB6 catalog \citep{Gr96}, as well as a
1.4~GHz measurement of 141$\pm$5~mJy from the NRAO VLA Sky Survey
\citep{Co98}. To constrain the dust emission, we used broadband fluxes from
the IRAS point source catalog v2.0 at 60 and 100~$\mu$m (IPAC 1986), from
Spitzer MIPS images at 70 and 160~$\mu$m \citep{Ue06}, and from the PACS and
SPIRE images presented in this paper at all five wavelengths. The data are
summarized in Table~\ref{irfluxes}. Since the source is extended, the
appropriate correction for extended source calibration was applied to the
SPIRE fluxes.
\begin{equation}
F_\nu{\rm [actual]} = \frac{K_{\rm 4E}}{K_{\rm 4P}}\frac{F_\nu{\rm [quoted]}}{K_{\rm color}}
\end{equation}
The correction factors are given in Table~\ref{irfluxes}. Note that the
definition of $K_{\rm color}$ is the inverse of the definition used in the
SPIRE observers manual, but agrees with the definition used by the other
instrument teams. We fitted a modified blackbody to the data, given by:
\[ F_\nu = C \nu^\beta B_\nu(T_{\rm dust}). \]
The resulting parameters were $T_{\rm dust} = 29.9\pm1.1$~K and $\beta =
2.12\pm0.12$. The value for $\beta$ is within the uncertainty equal to the
theoretically expected value $\beta = 2$ for optically thin dust. The color
corrections were based on this model. The actual fluxes and the best fit model
are shown in Fig.~\ref{sed}.

The diameter of the bar-like region is 100~arcsec which we used to constrain
the diameter of the ionized region. For the height of the torus we assumed
0.193~pc. If viewed exactly edge-on, this would equate to 33\arcsec\ at the
assumed distance of NGC 650.

\subsection{Dereddening the spectra}
\label{dered}

Before we could model NGC 650, we needed to correct the observed line ratios
for the effects of interstellar and circumstellar reddening. First we
determined the extinction coefficient c(H$\beta$) averaged over the nebula. We
did this by comparing the H$\beta$ flux derived from the optically thin
4.85~GHz radio flux to the observed H$\beta$ flux. To calculate the predicted
H$\beta$ flux, we used the following equation:
\begin{equation}
\frac{S_\nu}{F({\rm H}\beta)} = 2.51\times10^{10} \ T_{\rm e}^{0.53}
\nu_9^{-0.1} \, Y \hspace{2mm} \left[ \frac{\rm Jy}{\rm W\,m^{-2}} \right],
\end{equation}
taken from \citet{PottaschBook}. In this formula, $S_\nu$ is the observed
radio flux in Jy, $T_{\rm e}$ is the electron temperature averaged over the
ionized region, $\nu_9$ is the frequency of the radio observation in GHz, and
$Y$ is given by:
\begin{equation}
Y = 1 + \frac{n({\rm He^+})}{n({\rm H^+})} + 3.7 \, \frac{n({\rm He^{2+}})}{n({\rm H^+})}.
\end{equation}
Using the data from a preliminary Cloudy model of NGC 650, we found that $Y =
1.214$ and the average $T_{\rm e} = 11.8$~kK. Using these values yielded a
predicted value $\log F({\rm H}\beta) = -13.52$~W\,m$^{-2}$. When compared to
the observed value $\log F({\rm H}\beta) = -13.68$~W\,m$^{-2}$ \citep{Ka78}
this yielded $c({\rm H}\beta) = 0.16 \pm 0.04$~dex. Using the same procedure
on the observed 1.4~GHz flux yielded $c({\rm H}\beta) = 0.20 \pm 0.03$~dex.
Both values are in agreement within the uncertainty. We adopted the average of
the values: $c({\rm H}\beta) = 0.18 \pm 0.03$~dex.

Various determinations of the extinction derived from optical spectra using
the Balmer decrement method can be found in the literature. \citet{PT87}
report $c({\rm H}\beta) = 0.35$~dex in a spectrum centered on the high surface
brightness region in the south-west of the bar. \citet{Kw03} report $c({\rm
  H}\beta) = 0.08$~dex in the NGC 650A spectrum (centered on a low-density
region close to the central star) and $c({\rm H}\beta) = 0.21$~dex in the NGC
650B spectrum (also centered on the high surface brightness region in the
south-west of the bar). These data indicate that the extinction is varying
over the nebula. This is confirmed by the analysis of \citet{RL08} who
conclude that there are appreciable levels of clumping within the central bar.
They also report a gradual increase in extinction along the major axis of the
bar towards the north-east. They conclude that probably much of the dust
responsible for the extinction is located within the nebula itself, i.e.\ that
the nebula has an appreciable amount of internal extinction. This is
remarkable for a highly evolved nebula like NGC 650.

To deredden the UV data, we also needed to consider the shape of the
extinction curve, usually parameterized by the ratio of absolute to selective
extinction $R_V$. It is well known that this parameter differs substantially
for different sight lines through the interstellar medium (ISM) (e.g.\ Fig.~2
of \citealp{Fi99}) with values ranging between $R_V = 2.2$ and $5.8$. If the
nebula has an appreciable amount of internal extinction, this may result in an
extinction curve that is unlike a normal ISM extinction curve. We assumed that
we could use the parameterized extinction curve given by \citet{Fi99}. To do
the dereddening, we demanded that the dereddened line ratio of He\,{\sc ii}
$\lambda$1640/$\lambda$4686 equals the theoretical value of 6.35. Doing this
while keeping $R_V$ at the standard value of 3.1 yielded $c({\rm H}\beta) =
0.604$~dex, which implies $A_V = 1.27$~mag. Inspecting Fig.~5 of \citet{RL08}
showed that this value is not found anywhere in the bar, clearly excluding
this value for $R_V$. Hence we were forced to treat $R_V$ as a free parameter.
To determine the value of $R_V$, we combined the criterion above with the
requirement that the extinction equals $A_V = 0.7$~mag, an average of the data
in Fig.~5 of \citet{RL08} over the various IUE apertures. This yielded $R_V =
2.35$ and $c({\rm H}\beta) = 0.346$~dex. Such a low value for $R_V$ is
indicative of dust with a large amount of carbonaceous material and/or a
smaller than usual grain size distribution (e.g., due to the presence of
PAHs). This would be consistent with the presumed presence of very small
carbonaceous grains (possibly PAHs) in the clumps, in addition to the larger
grains. See Sect.~\ref{heating} for further discussion. The fact that we found
a higher $c({\rm H}\beta)$ than with the radio method is consistent with the
fact that there is internal extinction in the bar.

Given that the extinction in NGC 650 is very patchy and the optical spectra
cover only a small fraction of the bar, we used the Balmer decrement method
and derived a separate value of $c({\rm H}\beta)$ for each of the optical
spectra. Using $R_V = 2.35$ and requiring that the dereddened flux ratio
$F({\rm H}\alpha)$/$F({\rm H}\beta)$ equals the theoretical ratio 287 we found
$c({\rm H}\beta) = 0.065$~dex for the OPT-A spectrum and $c({\rm H}\beta) =
0.173$~dex for the OPT-B spectrum.

Given that the IRS apertures cover a large area of the bar region (at least
the LL aperture does) we dereddened the IRS spectra using the same extinction
parameters as the IUE data.

\subsection{Modeling results}
\label{model:results}

Since the nebula is reported to be carbon-rich \citep{KH96}, we assumed
graphite dust to be present in the nebula. Early modeling results indicated
that assuming the default Orion grain size distribution in Cloudy (which is an
approximation for $R_V = 5.5$ material) yields a higher grain temperature than
the observations allow, no matter how we varied the geometry. We therefore
assumed that the grains in NGC~650 are large and we approximated them with
single-sized grains with the radius as a free parameter. The best fit with the
observed SED was obtained by using grains with a radius of 0.15~$\mu$m.
Single-sized grains are not realistic for NGC~650, but apart of the observed
SED we have no additional information to constrain the size distribution. So
instead of introducing poorly constrained additional free parameters in the
size distribution, we decided to adopt a single-size approximation. The Cloudy
model yielded an average grain temperature of 31.5~K, in good agreement with
the observations. We assumed a constant density in the nebula and used the
350~$\mu$m flux as a stopping criterion for the model. The resulting model is
shown in Table~\ref{model:par}. It will be discussed in detail in
Sect.~\ref{discussion}. Our model is simultaneously ionization and density
bounded as the hydrogen ionization fraction is almost exactly 50\% at the
outer edge. This implies that the Cloudy model does not predict the presence
of a PDR and therefore also does not predict any
molecular emission. This is not in agreement with observations as NGC~650 has
been detected in H$_2$ \citep{ZG88,Kastner96,ML13} but not CO
\citep{HH89,Hu96}. This will be discussed further in Sect.~\ref{molecules}.

The fact that we find larger than normal grains in the ionized region seems
contradictory with our findings in Sect.~\ref{dered} where we mention the
presence of very small grains in the dense molecular knots. However, we point
out that very small grains (PAHs) can only survive in the dense knots where
they are shielded from the ionizing radiation. Including them in our modeling
would have been inappropriate since we only model the ionized gas.
Furthermore, the very small grains are expected to have a much smaller total
mass than the large grains, making their emission unimportant at wavelengths
$\geq 60~\mu$m. This is discussed further in Sect.~\ref{heating}.

\begin{table}
\caption{Parameters of the Cloudy model of NGC~650. $T_{\rm eff}$ and
  $L_{\ast}$ are the stellar temperature and luminosity, $r_{\rm in}$ and
  $r_{\rm out}$ are the inner and outer radius of the nebula, $n_{\rm H}$ is
  the hydrogen number density in the gas, $T_{\rm e}$ and $n_{\rm e}$ are the
  average electron temperature and density in the nebula, $\Gamma$ is the
  dust-to-gas mass ratio, m$_V$ is the unreddened {\it V}-band magnitude of
  the central star, $D$ is the distance, and $\epsilon$ is the elemental
  abundance by number relative to $\epsilon({\rm H}) \equiv 12$. Elemental
  abundances derived from only a single observed line are marked uncertain.
  Solar abundances were taken from \citet{Gr10}.}
\label{model:par}
\begin{tabular}{lr|lrr}
\hline
parameter & value & parameter & value & rel. to solar \\
\hline
$T_{\rm eff}$ (kK)            & 208.      & $\epsilon$(He)                & 11.04    & $+$0.11 \\
$L_{\ast}$ ($L_{\odot}$)      &  261.      & $\epsilon$(C)                & 8.94      & $+$0.51 \\
$r_{\rm in}$ (mpc)            & 93.        & $\epsilon$(N)                 &  8.20    & $+$0.37 \\
$r_{\rm out}$ (mpc)           & 274.       & $\epsilon$(O)                 &  8.62    & $-$0.07 \\
log($n_{\rm H})$ (cm$^{-3}$)  & 2.14       & $\epsilon$(Ne)                &  7.97    & $+$0.04 \\
$T_{\rm e}$ (kK)              & 12.13     & $\epsilon$(Mg)                 &  6.60\u  & $-$1.00\u \\
log($n_{\rm e})$ (cm$^{-3}$)  & 2.17       & $\epsilon$(Si)                &  6.56    & $-$0.95 \\
log($\Gamma$)                 & $-$2.22  & $\epsilon$(P)                &  5.41\u  & $+$0.00\u \\
m$_V$                         & 18.04    & $\epsilon$(S)                &  6.79    & $-$0.33 \\
$D$ (pc)                      &  1200.   & $\epsilon$(Cl)                &  4.92    & $-$0.58 \\
$\chi^2$                      & 5.15     & $\epsilon$(Ar)                &  6.37    & $-$0.03 \\
\hline
\end{tabular}
\end{table}

\section{Discussion}
\label{discussion}

In this Section we give a detailed discussion of the central star of NGC 650
and the physical properties of the nebula (with emphasis on the dust grains)
using the Cloudy model that we derived in Sect.~\ref{model}. The model input
parameters, as well as a few derived quantities and the overall $\chi^2$ of
the fit, have been presented in Table~\ref{model:par}.

\subsection{The central star temperature}
\label{star:temp}

With a temperature of 208~kK, the central star is clearly of a very high
excitation class. The low luminosity of 261~$L_\odot$ indicates that the star
is already well evolved on the cooling track and that all internal nuclear
reactions have ceased. Comparing these numbers to the evolutionary tracks
calculated by \citet{Bl95} yield the best fit to the track with a zero-age
main sequence (ZAMS) mass of 7~$M_\odot$ (core mass 0.940~$M_\odot$). The age
of the PN implied by the model track is roughly 9000~yr. When we combine the
expansion velocities given by \citet{Bryce96} (60~km\,s$^{-1}$ for the inner
lobes) with the angular diameter (170\arcsec) and distance (1200~pc), we find
a kinematical age of 8000~yr, in excellent agreement with the model track.
However, we also need to consider the uncertainties in this determination.
Central stars with a ZAMS mass of 7~$M_\odot$ are expected to undergo
hot-bottom burning on the AGB which efficiently converts carbon into nitrogen,
thus preventing the stars from becoming carbon-rich. However, this process
reduces in efficiency when the envelope mass is reduced, thus opening up the
possibly for more massive stars to become carbon-rich towards the end of the
thermally-pulsing AGB phase \citep{Fr98,He05}. A further consideration we need
to take into account is that hydrogen-deficient tracks yield higher
temperatures than hydrogen-rich ones for the same core mass \citep[and
  references therein]{WH06}. This is a concern because we used a hydrogen-rich
track while we know that the central star is hydrogen-deficient.
Unfortunately, only a few calculations exist for hydrogen-deficient central
stars and none of them model sufficiently massive central stars for our needs.
All we can say is that the \citet{VW94} hydrogen-deficient tracks indicate a
ZAMS mass larger than 2~$M_\odot$. We should also take the uncertainties in
stellar evolution modeling into account. See \citet{He08} for a review of the
uncertainties involved in modeling the AGB phase.

Other possible sources of uncertainty are the determination of the distance
and/or the stellar temperature. The uncertainty in the distance determination
is difficult to assess, but is likely substantial (see Sect.~\ref{intro}).
There are also concerns regarding the central star temperature. We see that
the stellar temperature we derived from nebular modeling is considerably
higher than the value of 140~kK derived from the observed stellar spectrum
\citep{NS95}. The stellar temperature seems quite well constrained given that
we see many different ionization stages in the spectra. However, the highest
ionization stages are not observed, most notably Ne$^{5+}$ and Ar$^{5+}$. The
[Ar\,{\sc vi}] 4.53~$\mu$m over [Ar\,{\sc v}] 13.1~$\mu$m line flux ratio, and
more importantly the [Ne\,{\sc vi}] 7.65~$\mu$m over [Ne\,{\sc v}] 14.3~$\mu$m
ratio, would have been strong constraints for the stellar temperature.
Unfortunately, the [Ar\,{\sc vi}] 4.53~$\mu$m line is outside the IRS
wavelength range and the [Ne\,{\sc vi}] 7.65~$\mu$m line has not been detected
due to the faintness of the line (note however that the non-detection is
consistent with our model).

We investigated the accuracy of our temperature determination by creating
additional fixed-temperature models with the stellar temperature and the inner
radius fixed while optimizing all remaining parameters. The stellar
temperature was fixed at values between $-0.1$ and $+0.1$~dex from the optimal
value of 208~kK. The inner radius was kept fixed at 93~mpc since that value
agrees well with the observed morphology. This is necessary since otherwise
the code would alter the inner radius in an attempt to keep the ionization
parameter constant and thus counteract the effect of changing the stellar
temperature. We searched for stellar temperatures such that the $\chi^2$ would
be 1 higher than the optimal value. This procedure gives us an estimate of the
1$\sigma$ uncertainty of the stellar temperature. The result is $T_{\rm eff} =
208^{+54}_{-32}$~kK. We found that the most important temperature sensitive
ratio is the [Ne\,{\sc v}] 14.3~$\mu$m over [Ne\,{\sc iii}] 15.6~$\mu$m ratio,
which is well constrained by the LL2 spectrum. We also checked that an
alternative Cloudy model with the central star temperature fixed to 140~kK can
be ruled out at the 7$\sigma$ level (keeping the inner radius fixed as
discussed above). The alternative model predicted a line flux ratio of 0.0265
compared to the observed value of 0.232 and 0.207 predicted by the optimal
Cloudy model. Using this procedure we derived a 3$\sigma$ lower limit for the
temperature of 165~kK.

One possibility we need to consider is that the nebular material is not in
photoionization equilibrium with the central star. This could be the case
because the central star is cooling down and also dropping in luminosity which
will cause the nebular gas to recombine. This process will take a finite time,
implying that the nebula will reflect the higher temperature and luminosity of
the central star at an earlier time. To derive a timescale for the reaction of
the gas we will look only at the recombination of a highly ionized ion. In
particular we will use Ne$^{4+}$ since that is the most highly ionized ion
that was detected and spectral lines from this ion are crucial to constrain
the stellar temperature. For gas with the parameters given in
Table~\ref{model:par} the total recombination coefficient is $3.15 \times
10^{-11}$~cm$^3$\,s$^{-1}$, which then yields a recombination timescale of
6.8~yr. Similarly we find 8.4~yr for Ar$^{4+}$ and 13.8~yr for O$^{3+}$. These
timescales are much shorter than the evolutionary timescale so that we can
safely assume that at least the most highly ionized ions are effectively in
equilibrium with the central star. We therefore feel confident that a stellar
temperature of 140~kK is ruled out by the Cloudy modeling.

As we already remarked in Sect.~\ref{intro}, the central star spectrum used by
\citet{NS95} to derive the central star parameters seems to have rather low
S/N ratio. This could cast doubt on their observation that the spectrum is
very similar to the spectrum of PG 1159, which is well studied \citep{We91}.
Furthermore, the optical spectrum of PG 1159 shows highly excited lines of
C\,{\sc iv}, N\,{\sc v}, O\,{\sc vi}, as well as He\,{\sc ii}. The CNO lines
will be crucial for determining the central star temperature. The 140~kK model
atmosphere shows that CNO is nearly fully in its helium-like state throughout
the atmosphere (see Fig.~4 of \citealp{We91}). Raising the stellar temperature
by moderate amounts would not change this as an enormous amount of energy is
needed to ionize CNO into its hydrogen-like state (viz. 392~eV for H-like
carbon, compared to 138~eV for He-like oxygen). This could suggest that the
optical spectrum becomes insensitive to temperature changes above 140~kK. We
will however not investigate this conjecture any further as it is well outside
the realm of this paper.

We could try to verify the stellar temperature by comparing the predicted
absolute {\it V}-band magnitude $M_V$ = 7.64~mag from the stellar atmosphere
model with the observed dereddened {\it V}-band magnitude $m_V$ = 17.00~mag
from \citet{KP98}. This requires knowledge of the distance of the star.
Unfortunately we cannot use the gravity distance of 1200~pc from \citet{KP98}
for this purpose as this is based on an assumed temperature of the central
star. Doing so would invert their analysis and by definition give back the
assumed value of 140~kK. So we are not able to carry out this analysis until
an accurate distance based on an independent method becomes available.

An additional constraint for the stellar temperature could be obtained from
the upper limit of the X-ray flux determined with Chandra \citep{Ka12}. We
used the absolutely calibrated central star spectra from our additional
fixed-temperature models that we described above and predicted the Chandra
count rate for each of those. Using this procedure we could determine a
3$\sigma$ upper limit for the central star temperature of 178~kK. In this
calculation we included the effects of internal extinction in the nebula ($A_V
= 0.1$~mag) but not interstellar extinction. The total extinction towards the
central star is unknown. The extinction measured in the OPT-A spectrum is $A_V
= 0.13$~mag. If we assume this value for the central star as well, this would
raise the 3$\sigma$ upper limit for the central star temperature to 183~kK. A
total extinction of $A_V = 0.2$~mag towards the central star would raise the
upper limit to 193~kK. In our photoionization modeling, such a temperature
would be allowed at the 1$\sigma$ level. It should be noted however that such
a central star has a very soft spectrum, with most of the photons in the 0.2
-- 0.3~keV range. In this range the calibration of Chandra is quite uncertain.

All in all we find that there are many uncertainties making a comparison with
theoretical tracks difficult, but we are confident that the progenitor of the
central star had a mass of at least 3~$M_\odot$. This lower limit is based on
the kinematical age of the nebula and the nebular abundance pattern (see
Sect.~\ref{abundances}) both of which should be relatively robust
determinations.

\subsection{The AGB shell}

No spherical halo has ever been observed beyond the lobes of NGC 650 and we
also do not detect a spherical dusty AGB halo, as has been the case for e.g.
NGC~6720. NGC~650 has a PG 1159 central star. The origin of the PG 1159 stars
can possibly be explained by a very late thermal pulse initiating the
born-again post-AGB scenario \citep{Alt08}. The born-again scenario was first
described by \citet{Sch79} and \citet{Iben83}. If so, the PG 1159 stars would
go through the post-AGB evolution up to three times \citep{Ha05,vH07}. Given
the unusually high expansion velocity of the nebula, it is possible that the
ejecta from the second post-AGB loop have overtaken the original PN by now.

\subsection{Elemental abundances}
\label{abundances}

Our modeling confirms that the gas in NGC 650 is carbon rich, with C/O $=
2.1$. This is lower than the value $2.96 \pm 0.50$ derived by \citet{KH96}.
The latter value is based solely on the \ms{C\,{\sc iii}} $\lambda1909$ line,
while we include the \ms{C\,{\sc ii}} $\lambda2326$ and C\,{\sc iv}
$\lambda1549$ lines as well. The main reason for the discrepancy is the
assumed H$\beta$ flux corresponding to the IUE aperture, as was already
discussed in Sect.~\ref{sec:iue}. The H$\beta$ flux we assume is about a
factor 2 higher, which lowers the strength of the carbon lines by the same
amount. Other factors that contribute to the change in the C/O ratio are the
fact that we average the flux over six apertures (this raises the \ms{C\,{\sc
    iii}} $\lambda1909$ flux by some 10\%) but more importantly that we do the
dereddening using an $R_V = 2.35$ law (this raises the dereddened \ms{C\,{\sc
    iii}} $\lambda1909$ flux by nearly 20\%). The fact that the gas is carbon
rich justifies our choice to use carbonaceous grains.

The abundances are typical for a type IIa PN \citep{M00} with He/H = 0.11 and
log(N/O) = $-0.42$, using the classification scheme of \citet{FAM87}. This
would make the progenitor star a fairly massive galactic disk object. This is
consistent with the high stellar temperature that we derived.

\subsection{Line contribution to the broadband fluxes}
\label{sec:syn}

\begin{figure}
\begin{center}
\includegraphics[width=\columnwidth]{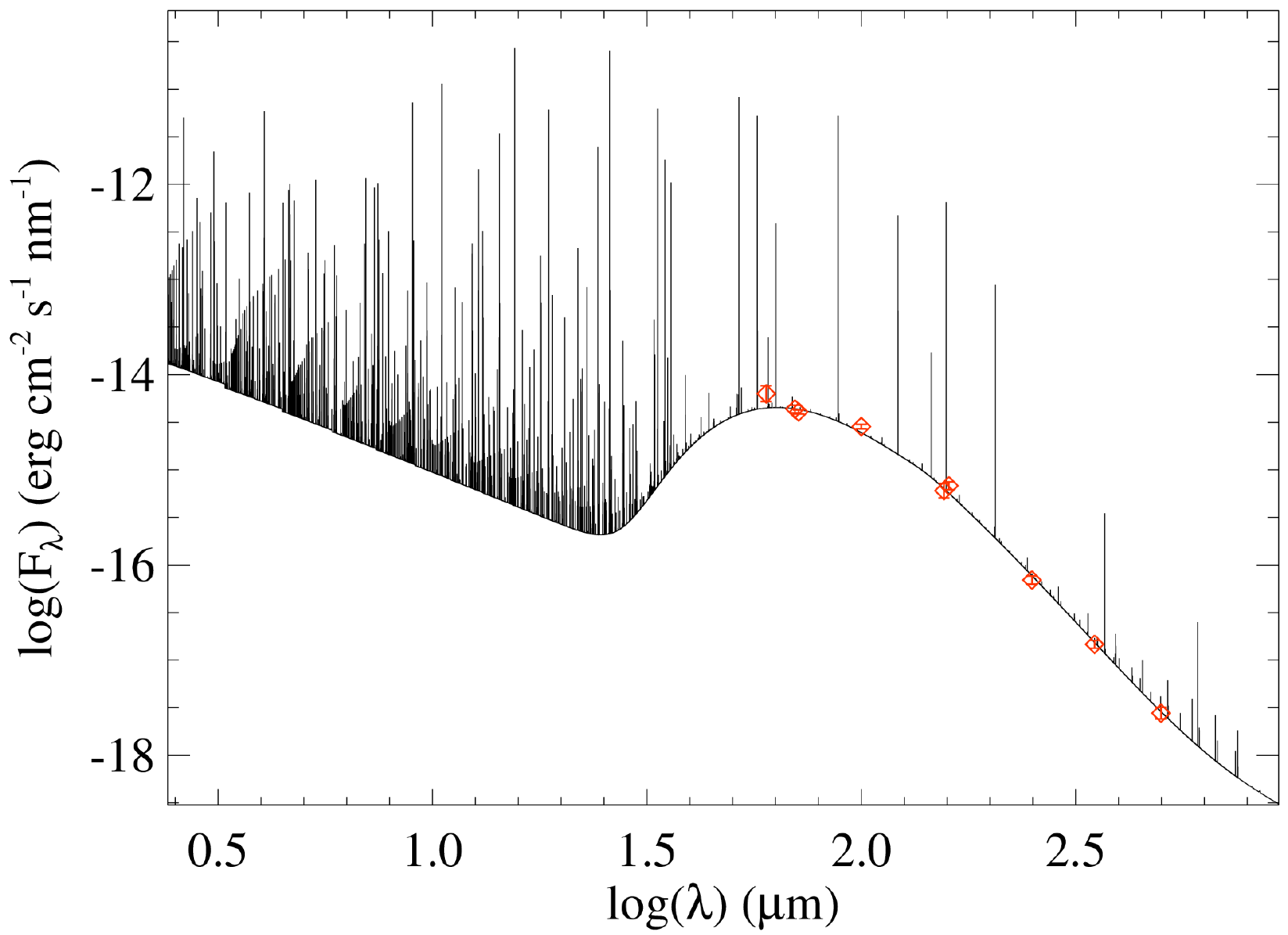} 
\caption{The infrared spectrum computed by Cloudy. The far-IR photometry from
  Table~\ref{irfluxes} (shown as red diamonds) is also included to demonstrate
  the quality of the fit.}
\label{cldspec}
\end{center}
\end{figure}

The various broadband fluxes that we presented in this paper will not only
contain thermal emission from dust grains, but also various emission lines. We
can use the Cloudy model to assess the line contribution in each of the
photometric bands. We do this by taking the absolutely calibrated model
spectrum (shown in Fig.~\ref{cldspec}) and fold this with the various
passbands to obtain synthetic photometry. This model spectrum contains
continuum and line emission, but these two components can be separated
allowing us to get detailed information about the line contribution in each
passband. We did not include the incident spectrum in the model as this
includes the cosmic microwave background (CMB) emission. This component would
raise the SPIRE 500 $\mu$m flux noticeably and would be inconsistent with
observations since the CMB emission was removed in the background subtraction
procedure (it is uniform over the image). For passbands at shorter wavelengths
the omission of the central star spectrum has no effect since we checked that
it is negligible in all the bands we consider. In Table~\ref{synthetic} we
present the synthetic fluxes we obtained combined with all emission lines
contributing more than 3\% of the inband flux. Note that the Cloudy model does
not model the high density clumps where the H$_2$ resides (see
Sect.~\ref{molecules}). Hence the H$_2$ lines are not included in the
synthetic photometry. We expect that this has little impact on the predictions
as these lines are weak in this PN (see Sect~\ref{molecules}).

\begin{table*}
\caption{Synthetic photometry of NGC 650. The first column gives the
  photometric band. The second column gives the synthetic flux derived from
  the Cloudy model. The third column gives the line contribution (lc) in
  percent. The fourth column gives the quoted observed flux. The fifth, sixth
  and seventh column give individual lines contributing at least 3\% to the
  inband flux. Laboratory wavelengths for the emission lines are given in
  vacuum.}
\label{synthetic}
\begin{tabular}{lrrrlrr}
\hline
band & synth.\ flux & lc & obs.\ flux & line id. & $\lambda_{\rm lab}$ & lc \\
     & Jy           & \% & Jy         &          & $\mu$m             & \% \\
\hline
IRAC    3.6 & 0.0324 & 15.4 & 0.0359 & H\,{\sc i} 8--5   &  3.740556 &  6.01 \\
            &        &      &        & H\,{\sc i} 9--5   &  3.296992 &  3.46 \\
IRAC    4.5 & 0.0555 & 48.9 & 0.0708 & H\,{\sc i} 5--4   &  4.052262 & 22.5  \\
            &        &      &        & H\,{\sc i} 7--5   &  4.653778 &  5.74 \\
            &        &      &        & [K\,{\sc iii}]    &  4.6180   &  5.00 \\
            &        &      &        & He\,{\sc ii} 8--7 &  4.763508 &  3.81 \\
            &        &      &        & [Mg\,{\sc iv}]    &  4.4867   &  3.81 \\
IRAC    5.7 & 0.0392 & 25.4 & 0.0359 & [Fe\,{\sc ii}]    &  5.340263 & 11.1  \\
            &        &      &        & H\,{\sc i} 9--6   &  5.908213 &  3.03 \\
IRAC    7.9 & 0.1036 & 70.3 & 0.1067 & [Ar\,{\sc iii}]   &  8.99138  & 48.8  \\
            &        &      &        & H\,{\sc i} 6--5   &  7.459858 &  5.25 \\
            &        &      &        & [Ar\,{\sc ii}]    &  6.985274 &  4.87 \\
            &        &      &        & [Na\,{\sc iii}]   &  7.3178   &  4.36 \\
WISE    3.4 & 0.0342 & 19.7 & ---    & H\,{\sc i} 8--5   &  3.740556 &  4.46 \\
            &        &      &        & He\,{\sc ii} 7--6 &  3.091693 &  4.40 \\
WISE    4.6 & 0.0453 & 37.3 & ---    & H\,{\sc i} 5--4   &  4.052262 &  7.28 \\
            &        &      &        & H\,{\sc i} 7--5   &  4.653778 &  6.65 \\
            &        &      &        & [K\,{\sc iii}]    &  4.6180   &  5.91 \\
            &        &      &        & He\,{\sc ii} 8--7 &  4.763508 &  4.51 \\
            &        &      &        & [Mg\,{\sc iv}]    &  4.4867   &  3.83 \\
WISE   11.6 &  0.353 & 90.7 & 0.62   & [Ne\,{\sc iii}]   & 15.55505  & 48.7  \\
            &        &      &        & [S\,{\sc iv}]     & 10.51049  & 15.9  \\
            &        &      &        & [Ne\,{\sc v}]     & 14.32168  &  9.47 \\
            &        &      &        & [Ar\,{\sc iii}]   &  8.99138  &  8.31 \\
            &        &      &        & [Ne\,{\sc ii}]    & 12.813548 &  3.05 \\
WISE   22.1 &  0.483 & 92.2 & 3.92   & [O\,{\sc iv}]     & 25.8903   & 61.8  \\
            &        &      &        & [Ne\,{\sc v}]     & 24.3175   & 25.8  \\
            &        &      &        & [Ar\,{\sc iii}]   & 21.8291   &  3.57 \\
IRAS   12   &  0.210 & 83.2 & 0.28   & [S\,{\sc iv}]     & 10.51049  & 36.1  \\
            &        &      &        & [Ar\,{\sc iii}]   &  8.99138  & 18.3  \\
            &        &      &        & [Ne\,{\sc v}]     & 14.32168  & 16.2  \\
            &        &      &        & [Ne\,{\sc ii}]    & 12.813548 &  5.93 \\
IRAS   25   &  1.383 & 96.5 & 2.79   & [O\,{\sc iv}]     & 25.8903   & 79.7  \\
            &        &      &        & [S\,{\sc iii}]    & 18.71303  &  8.78 \\
            &        &      &        & [Ne\,{\sc v}]     & 24.3175   &  7.75 \\
IRAS   60   &  6.96  & 32.3 & 6.80   & [O\,{\sc iii}]    & 51.8145   & 16.5  \\
            &        &      &        & [N\,{\sc iii}]    & 57.34     & 14.1  \\
IRAS   100  & 11.26  & 27.9 & 9.29   & [O\,{\sc iii}]    & 88.356    & 25.2  \\
MIPS   23.7 &  1.519 & 97.4 & 4.51   & [O\,{\sc iv}]     & 25.8903   & 84.5  \\
            &        &      &        & [Ne\,{\sc v}]     & 24.3175   & 12.7  \\
MIPS   71.4 &  7.55  & 14.7 & 6.04   & [O\,{\sc iii}]    & 88.356    &  6.31 \\
            &        &      &        & [N\,{\sc iii}]    & 57.34     &  5.89 \\
MIPS  155.9 &  7.28  & 30.6 & 4.83   & [C\,{\sc ii}]     &157.68     & 29.8  \\
PACS   70   &  7.64  &  9.1 & 6.95   & [O\,{\sc iii}]    & 88.356    &  3.86 \\
PACS   100  & 12.10  & 33.3 & ---    & [O\,{\sc iii}]    & 88.356    & 29.7  \\
PACS   160  &  6.54  & 22.7 & 6.05   & [C\,{\sc ii}]     &157.68     & 20.2  \\
SPIRE  250E &  1.705 &  2.4 & 1.50   & ---               & ---       & ---   \\
SPIRE  350E &  0.655 &  1.0 & 0.62   & ---               & ---       & ---   \\
SPIRE  500E &  0.247 &  0.4 & 0.24   & ---               & ---       & ---   \\
\hline
\end{tabular}
\end{table*}

In addition to the line emission, there is also continuum emission
contributing to the inband flux. The continuum emission can be characterized
as follows. In the short wavelength bands ($\lambda < 12~\mu$m) nebular
free-free and free-bound emission will contribute to the inband flux, while
the dust emission is negligible. At longer wavelengths the dust continuum will
be the dominant continuum contribution. It peaks around 80~$\mu$m and then
starts to drop again. Around 500~$\mu$m nebular free-free emission will become
noticeable again and longward of 1~mm, free-free emission will be the dominant
continuum source. Hence there is a small contribution of free-free emission to
the red wing of the SPIRE 500 band. This explains why this point is slightly
above the SED fit in Fig.~\ref{sed}. In the mid-IR range (12 to 25~$\mu$m) the
continuum is predicted to be weak so that the inband flux is almost completely
dominated by line emission. This could make synthetic photometry less accurate
if narrow features exist in the passband (e.g., due to incomplete
understanding of the transmission function or Doppler shifts in the lines).

Comparing the synthetic IRAC fluxes with the observed values (3.6~$\mu$m:
0.0359~Jy, 4.5~$\mu$m: 0.0708~Jy, 5.7~$\mu$m: 0.0359~Jy, 7.9~$\mu$m:
0.1067~Jy; \citealp{Hora04}) shows that they are in excellent agreement, with
the 5.7~$\mu$m band showing the largest discrepancy (around 25\%). The other 3
bands all agree within 10\%. Note that the IRAC fluxes were not used to
constrain the Cloudy model, so these present an independent validation of the
model. The long wavelength photometry ($\lambda > 60~\mu$m) is also in good
agreement with observations, even though the Cloudy model is systematically
slightly over-predicting the fluxes. This is because the observed fluxes in
this wavelength range were used to constrain the dust continuum in the Cloudy
model, but were not corrected for line emission. As a result the dust-to-gas
mass ratio will be somewhat too large. We estimate this error to be around
10\%. The IRAS 25 and MIPS 23.7~$\mu$m synthetic fluxes show the biggest
discrepancy with observations (2.79 and 4.51~Jy, respectively,
\citealp{Ue06}). This will be discussed further in Sect.~\ref{heating}.

\subsection{Properties of the Dust Grains}

Using the Cloudy model, we derived that the mass of the central bar is
0.23~$M_\odot$ (but note that our model does not explicitly model the high
density clumps in the bar, it assumes constant density). Using the dust-to-gas
mass ratio $\Gamma = 6.0 \times 10^{-3}$ that resulted from the Cloudy model,
we find that the dust mass in the bar of NGC~650 is $1.4 \times
10^{-3}$~$M_\odot$. The main uncertainty in this number will come from the
uncertainty in the distance determination.

In Sect.~\ref{model:results} we find that the dust grains in the ionized
region of NGC 650 are larger than typical grains in the ISM. This has been
seen in other PNe as well, e.g. NGC 6445 \citep{vH00}. Several explanations
could be proposed for this observation. The first explanation is that the
smallest grains have been destroyed in the circumstellar environment, either
through shocks or more gradual processes like thermal sputtering in the
ionized region. This explanation seems unlikely as grain destruction would
return fourth-row elements like calcium and iron to the gas phase. These
elements have very prominent emission lines in the optical region if they
would be at solar abundance. Such lines are not seen in the optical spectrum.
This is also confirmed by the analysis of \citet{St99} who find that there is
a considerable spread in the dust content of planetary nebulae, but that there
is no evidence for a decrease in the dust-to-gas mass ratio as the planetary
nebulae evolve. They conclude that the results suggest that the timescale for
destruction of dust grains in planetary nebulae is larger than their lifetime.

The second explanation could be that the grains have been processed in the
circumstellar environment. Since the density in a typical planetary nebula is
too low for such processes, this would require the presence of a circumstellar
disk. This process is known to be at work in some evolved objects like e.g.
the Red Rectangle \citep{Ju97}, M1--92 \citep{Mu10}, and other post-AGB
objects \citep{Gi11}. However, currently no evidence is known for the presence
of such a disk in NGC~650. It is also not clear how the grains would be
transported from the circumstellar disk to the torus, or if the circumstellar
disk could have evolved into the torus.

The third explanation could be that the large grains were already formed
during the AGB stage. The evidence for this is currently mounting, both for
oxygen-rich AGB stars \citep{No12,Gr12} and carbon stars \citep{Me07}. It has
also been seen in post-AGB objects (e.g., \citealp{Ue01}). This seems the most
plausible explanation and would imply that planetary nebulae inject mostly
larger than average grains into the ISM.

\subsection{Grain Heating}
\label{heating}

The temperature maps of NGC~650 are presented in Figs.~\ref{figtemp} and
\ref{figtempii}. In the temperature map of Fig.~\ref{figtempii} it is clear
that the hottest dust grains (as shown by the contours at 31.5~K and higher)
are only present in the low-density area along the polar axis of the torus.
The lower-temperature contours (30.5~K and lower) on the other hand are more
or less spherical and centered on the central star. The dust grains are
primarily heated by UV photons, either continuum photons emitted by the
central star, or diffuse emission from the gas (predominantly Ly$\alpha$
photons). For a discussion see \citet{vH00}. The morphology of the dust
temperature distribution suggests that the hottest dust grains are heated in
part by a radiation component that is readily absorbed by the torus, while the
radiation that is heating the cooler grains is clearly not absorbed by the
torus as is evident from the spherical contours. An alternative explanation
could be that the hottest grains are heated in part by shocks of a wind
impacting on the inside of the torus. However, in view of the highly evolved
status of the central star, the presence of a strong wind that would be needed
to create such a shock is not expected. The fact that NGC~650 is not a source
of diffuse X-ray emission \citep{Ka12} also argues against the presence of a
strong shock.

\begin{figure}
\begin{center}
\includegraphics[width=\columnwidth]{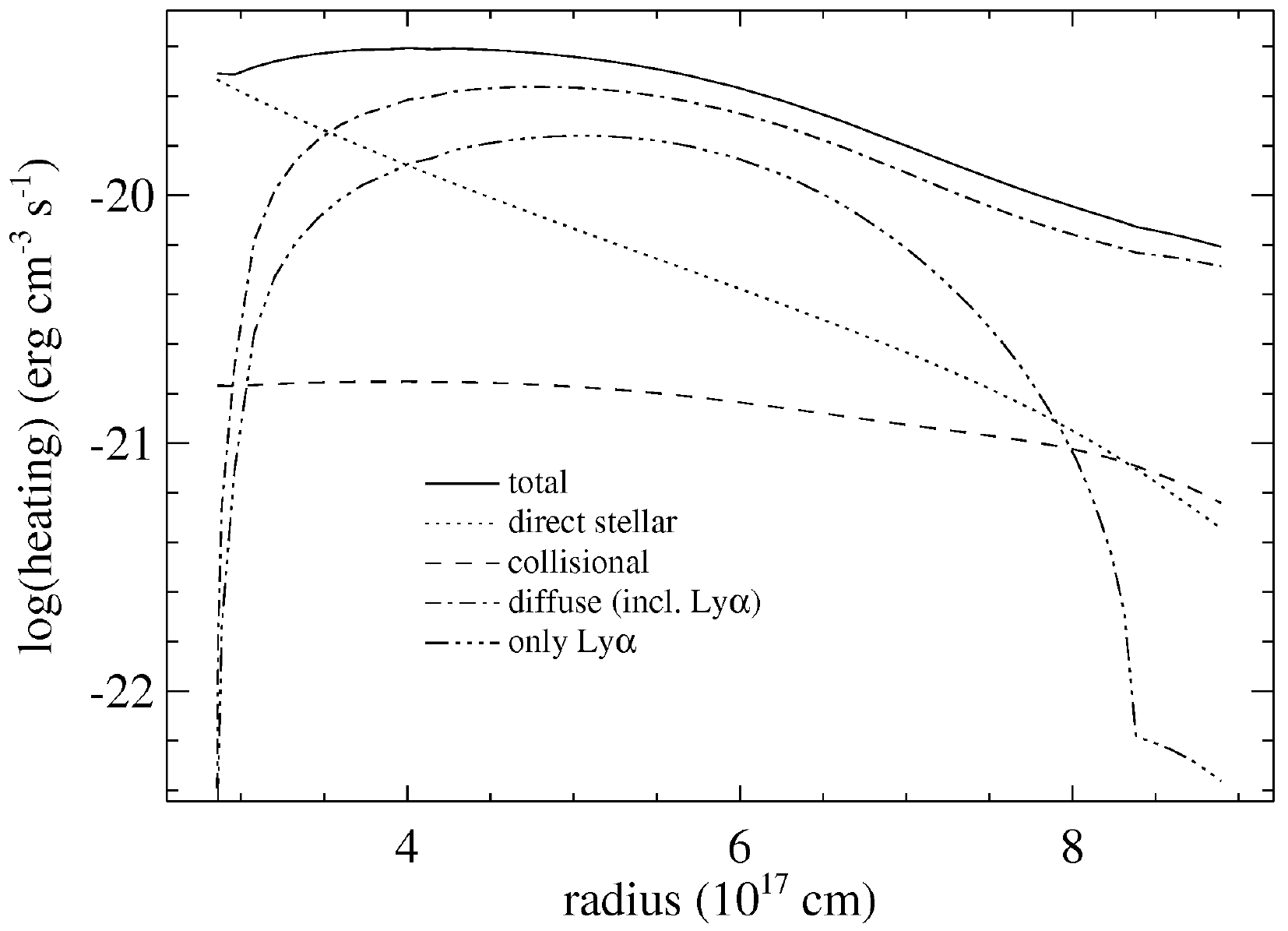} 
\caption{Various components contributing to the grain heating as a function of
  radius from the central star.}
\label{grheating}
\end{center}
\end{figure}

We can analyze the grain heating further using the Cloudy model. In
Fig.~\ref{grheating} we show the various components contributing to the
heating of the grains as a function of depth into the torus. These components
are: heating by incident radiation emitted directly by the central star,
collisional heating by the gas, and heating by diffuse emission from the
ionized gas. The latter component is dominated by Ly$\alpha$ photons which
contribute more than 50\% of the total diffuse heating. We have therefore
shown this component separately in Fig.~\ref{grheating} (note that the curve
labeled ``diffuse'' already includes the Ly$\alpha$ heating). This figure
shows that at the inner face incident starlight is an important heating
source. At greater depth the heating is quickly taken over by diffuse photons
and this remains so all the way up to the outer edge of the model. At this
point the reader should be aware of two shortcomings of the Cloudy model. The
first is that Ly$\alpha$ photons are heavily scattered by hydrogen atoms (see
also \citealp{vH00}). The photons make a random walk through the gas until
they are finally absorbed by a grain or some other background opacity. During
this random walk the photons can travel a considerable distance, which is not
modeled correctly in Cloudy. This implies that the distribution of Ly$\alpha$
photons will be flatter than shown in Fig.~\ref{grheating} and especially near
the inner face the Ly$\alpha$ density is expected to be higher than shown in
the plot. Consequently the fact that the total grain heating peaks at a finite
depth into the model is likely an artifact resulting from this deficiency.

The second shortcoming in the Cloudy model is that it assumes the gas to be
smooth, while in actual fact it has high density clumps that are very
efficient at absorbing the incident UV radiation. We know this because H$_2$
molecules reside in these clumps (see Sect.~\ref{molecules}). Depending on the
covering factor of these clumps, this may imply that the heating by incident
radiation drops off more quickly than suggested by the model. With these
shortcomings in mind we can say that, at least qualitatively, the Cloudy model
confirms the view that we derived from the analysis of the temperature map:
inside the hole of the torus and some shallow depth into the torus, incident
radiation from the central star will contribute significantly to the heating
of the grains. However, this component is quickly absorbed away, leaving only
diffuse photons as a heating agent. Note that Ly$\alpha$ photons are produced
everywhere in the ionized gas and can even be scattered into the dense clumps.
This explains the smooth, near spherical component in the grain heating. Note
that Ly$\alpha$ photons heating the grains in the clumps is not a
contradiction with the presence of H$_2$: Ly$\alpha$ photons cannot destroy
H$_2$ in the ground state (e.g., \citealp{Dr00}).

\citet{Ue06} presented a two-component fit to the dust emission including a
hot component of 140~K. This temperature was constrained by the IRAC
7.9~$\mu$m flux measurement of 0.107~Jy \citep{Hora04} and the MIPS
23.7~$\mu$m flux measurement of 4.51~Jy \citep{Ue06}. Looking at the synthetic
photometry in Table~\ref{synthetic} we can see that the measured IRAC flux is
in excellent agreement with the Cloudy model. We therefore feel confident that
no hot dust is contributing to the IRAC flux. The observed IRAS 12~$\mu$m flux
(0.28~Jy in the point source catalog -- PSC, and 0.255~Jy in the faint source
catalog -- FSC) is slightly higher than predicted by the model. This value is
close to the detection limit of IRAS, so to double-check this result, we also
measured the flux in the WISE 11.6~$\mu$m image. This yielded 0.62~Jy, also
higher than predicted by Cloudy. Hence hot dust could be contributing 0.1 --
0.2~Jy at 12~$\mu$m, but certainly not $\approx 1.5$~Jy as was predicted by
\citet{Ue06}. So the presence of a 140~K component with the strength suggested
by \citet{Ue06} can be ruled out. At longer wavelengths there is a rather
large discrepancy between the observed and synthetic flux in the MIPS 23.7 and
IRAS 25~$\mu$m bands (the observed values are: MIPS 4.51~Jy, IRAS PSC:
2.79~Jy, IRAS FSC: 3.14~Jy). The model predicts that the inband flux should be
dominated by [O\,{\sc iv}] emission. This is consistent with the fact the
nebula is clearly narrower in the MIPS 23.7~$\mu$m band \citep{Ue06}
(ionization stratification limits the [O\,{\sc iv}] emission to the inner
regions of the PN). However, the discrepancy with the observations cannot be
explained by problems with the model line flux since the [O\,{\sc iv}] flux is
constrained by observations and is fitted very well. The big discrepancy
between the observed MIPS 23.7 and IRAS 25~$\mu$m fluxes could indicate that
there are problems with the data reduction, e.g. due to an imperfect
background subtraction. This is because the synthetic fluxes in these bands
are nearly equal indicating that there is no known reason for the discrepancy
between the observed fluxes. It is also clear that there is excess continuum
emission that is not accounted for in the model (see Fig.~\ref{irsfig}).
Possible candidates would be stochastically heated very small grains (possibly
PAHs) and/or a separate (hotter) dust component.

\begin{figure}
\begin{center}
\includegraphics[width=\columnwidth]{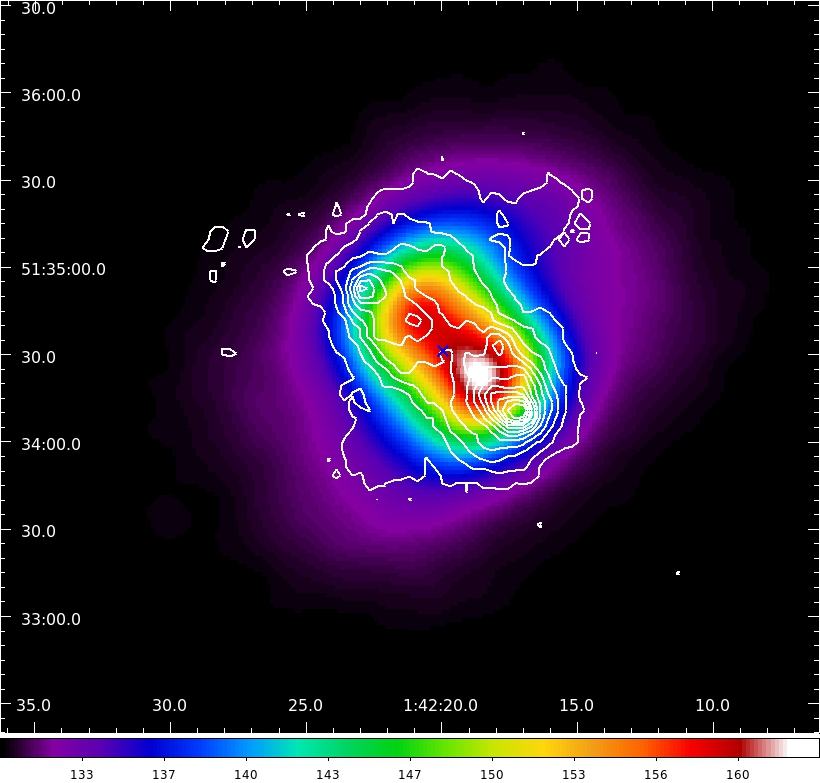} 
\caption{The WISE 22.1~$\mu$m image of NGC~650. The image is normalized in the
  standard DN units (1~DN = $5.2269\times10^{-5}$~Jy/pixel). The contours are
  taken from the PACS 70~$\mu$m image. The blue cross marks the position of
  the central star.}
\label{wise}
\end{center}
\end{figure}

To investigate this further we have measured the flux in the WISE 22.1~$\mu$m
band. This band has the advantage that the [O\,{\sc iv}] emission contributes
much less to the inband flux since the line is in the wings of the passband.
This is evident from the fact that synthetic WISE 22.1~$\mu$m flux is much
lower than the synthetic IRAS 25 and MIPS 23.7~$\mu$m flux. The observed value
for the WISE 22.1~$\mu$m flux is 3.92~Jy, in good agreement with the MIPS
23.7~$\mu$m flux. This indicates that the continuum dust emission that is not
accounted for in the Cloudy model must be the dominant contribution to the
WISE 22.1~$\mu$m flux. \citet{Ue06} hinted that the hot dust component could
be similar to the debris disk that was reported by \citet{Su07} in NGC 7293.
However, the WISE 22.1~$\mu$m image shown in Fig.~\ref{wise} shows a similar
morphology compared to the other images, only narrower, which makes such an
explanation unlikely. The peak of the emission clearly coincides with the
inner edge of the bright region in the SW of the bar and not the central star.
Since the continuum emission dominates the WISE passband, this is strong
evidence against a disk since that would have produced a strong point-like
peak at the location of the central star. Hence stochastically heated very
small grains in the dense clumps seem more likely. Such grains cannot survive
in ionized gas and could therefore only exist in the dense clumps where they
are shielded from the ionizing radiation. This implies that adding them to the
Cloudy model would not have made any difference since the Cloudy model stops
at the outer edge of the ionized region. The fact that the MIPS 23.7~$\mu$m
image appears narrower than the longer wavelength images can likely be
attributed to the fact that stochastic heating is less efficient in the outer
reaches of the bar due to extinction effects. The fact that the peak of the
WISE 22.1~$\mu$m emission is shifted towards the central star w.r.t. the peak
in the PACS 70~$\mu$m emission is also consistent with this. The UV photons
responsible for the stochastic heating emitted by the central star are readily
absorbed in the torus. The presence of very small grains would also be
consistent with the low $R_V$ value that we derived in Sect.~\ref{dered}. Most
likely the very small grains would have a much lower mass than the large
grains, implying that their emission would contribute very little in the PACS
and SPIRE bands. However, further research will be needed to settle this
point.

\subsection{Molecules in NGC 650}
\label{molecules}

Molecules are known to be present in NGC 650. More in particular, H$_2$ has
been detected \citep{ZG88,Kastner96,ML13}. By inspecting Fig.~8 of
\citet{ML13} it is clear that the H$_2$ emission has a very clumpy morphology.
This is very different from young massive PNe were the H$_2$ can be found
surrounding the ionized region as an unbroken layer (i.e., more like a classic
PDR, see e.g., \citealp{La00}).

Some authors \citep{Hora04,RL08} have argued for the existence of a PDR
surrounding the ionized region of NGC~650, similar to young PNe like e.g.,
NGC~7027. In both papers the argumentation is based on the IRAC images which
they assume to contain typical PDR emission, at least to some significant
level. More in particular, both authors observe that the IRAC 7.9~$\mu$m image
appears more extended than the shorter wavelength images. \citet{Hora04}
assume that this band contains H$_2$ emission while \citet{RL08} assume that
it contains PAH bands. We have analyzed the contributions to the IRAC bands in
Sect.~\ref{sec:syn} and concluded that the Cloudy model gives a very good
prediction for the inband flux despite the fact that it does not contain H$_2$
or PAH emission from a PDR. This leaves little room for additional flux
components. It is clear that H$_2$ emission will be contributing at some level
though, and possibly also PAH features. The H$_2$ contribution is likely to be
small, as will be discussed in more detail below. The PAH 6.2, 7.7, 8.6 and
11.2~$\mu$m features are not visible in the IRS spectra and therefore PAHs are
less likely to contribute much to the IRAC 7.9~$\mu$m band. However, it is
possible that the SL aperture happened to miss the dense clumps that could
contain the PAHs. The fact that we assume that we see stochastically heated
very small grains in the MIPS 23.7 and WISE 22.1~$\mu$m images does not
necessarily imply that significant PAH emission should be present in the IRAC
passbands. This depends on the highest temperature that the PAHs reach during
the temperature spikes. But even if PAHs were contributing significantly, it
can be clearly seen in the top panel of Fig.~1 of \citet{RL08} that the IRAC
7.9~$\mu$m emission is completely embedded in the H$\alpha$ emission from the
ionized region, clearly ruling out a PDR surrounding the ionized region of the
torus.

Therefore in this PN the H$_2$ only resides inside dense clumps that are
embedded in the ionized gas. This situation is reminiscent of other highly
evolved PNe like the Helix nebula (NGC 7293) and the Ring nebula (NGC 6720).
We therefore need to consider the conjecture that the same mechanism is
responsible for the formation of these clumps. The fact that the central star
is already well evolved on the cooling track indicates that the nebular gas is
either going through a recombination phase, or has gone through such a phase
in the recent past. In the latter case the nebula should have started to
ionize again due to expansion. In \citet{vH10} we proposed that the knots in
NGC 6720 formed during the recombination phase due to an instability caused by
rapid cooling of the gas. The fast cooling results in gas that is still
ionized (and thus produces abundant recombination radiation) but has very
little thermal pressure support. The radiation pressure of the recombination
radiation on the dust and/or gas could then cause the medium to become
unstable and fragment into many globules. The high density in the knots makes
it possible to re-form H$_2$ molecules on a suitable timescale.

The PN NGC~6781 is a bipolar nebula seen nearly pole-on that seems to have a
similar morphology to NGC 650 as it also possesses high-density clumps that
may be responsible for much of the molecular emission \citep{Ph11}. However,
in this nebula the H$_2$ emission lines detected in the IRS spectra are much
stronger than in NGC 650 and the emission extends to distances up to
60\arcsec\ beyond the rim. Further research will be needed to determine
whether this PN really is in a similar evolutionary state.

The IRS spectrum of NGC 650 is dominated by high-excitation atomic
fine-structure lines, as is typical for a hot PN. No H$_2$ emission lines have
been detected even though the (0,0) S(7) through (0,0) S(0) lines are covered
by the IRS wavelength range. H$_2$ lines are typically quite weak compared to
atomic fine structure lines, so this may be expected. \citet{ZG88} observed
NGC 650 using an aperture with a diameter of 19.6~arcsec on 8 positions. They
report detections of the (1,0) S(1) line on 3 positions (near the bright
region in the SW) between $7\times10^{-20}$ and $17\times10^{-20}$
W\,m$^{-2}$\,arcsec$^{-2}$ averaged over the aperture. Using these numbers we
estimate that this corresponds to a flux of roughly $10^{-16}$~W\,m$^{-2}$ in
the area covered by the LL aperture, though that number has a considerable
uncertainty. No other lines of H$_2$ have ever been measured, so the
excitation of the molecule is unknown. One could assume that the (0,0) S(5)
line is the strongest in this series \citep{Ph11} and comparable in strength
to (1,0) S(1). The noise in the SL2 spectrum (which covers the (0,0) S(7)
through (0,0) S(4) lines) is $3.8\times10^{-15}$~W\,m$^{-2}$\,$\mu$m$^{-1}$.
Using a FWHM of 0.2~$\mu$m for the line width, this is consistent with the
non-detections of the H$_2$ lines. The noise is much better in the SL1
spectrum ($1.2\times10^{-16}$~W\,m$^{-2}$\,$\mu$m$^{-1}$), but the lower lines
of the series are too weak to be detectable despite this. This makes it
unlikely that the H$_2$ lines contribute significantly to the IRAC 7.9~$\mu$m
band.

NGC~650 has not been detected in CO \citep{HH89,Hu96} in regions where the
H$_2$ emission is strongest, unlike NGC 7293 and NGC 6720. In PDR modeling it
is well known that CO forms at greater optical depths than H$_2$. In fact the
UV shielding provided by the H$_2$ molecules is needed to enable CO formation.
This could indicate that either the optical depth of the clumps in NGC 650 is
insufficient to reach the level of UV shielding needed, or that the H$_2$
abundance is too low and these molecules provide insufficient shielding. The
latter could then either have been caused by dissociation of existing
molecules or the fact there was insufficient time to re-form the necessary
amounts of H$_2$. Further research will be needed to investigate this.

\section{Conclusions}
\label{conclusions}

In this paper we presented new Herschel PACS and SPIRE images of NGC 650. We
used these images to derive a temperature map of the dust. We also constructed
a photoionization model using the spectral synthesis code Cloudy. To constrain
this model, we used the PACS and SPIRE fluxes and combined these with hitherto
unpublished IUE and Spitzer IRS spectra as well as various other data from the
literature. The temperature map combined with the photoionization model were
used to study various aspects of the central star, the nebula, and in
particular the dust grains in the nebula. These are the main conclusions.

\begin{itemize}
\item
The central star is hydrogen deficient and of PG 1159 (E) type. The
photoionization model shows that the central star is well evolved on the
cooling track. It yields an unusually high central star temperature of
$208^{+54}_{-32}$~kK and a luminosity of 261~$L_{\odot}$ assuming a distance
of 1200~pc. The central star temperature we derived is significantly higher
than the value of 140~kK that was derived from comparing the central star
spectrum with PG 1159. Further analysis yielded a 3$\sigma$ lower limit for
the stellar temperature of 165~kK (based on nebular modeling) and a 3$\sigma$
upper limit of 178~kK (based on a non-detection by Chandra). The upper limit
is consistent with the nebular model at the 1$\sigma$ level. The upper limit
would be higher if the extinction towards the central star is more than $A_V =
0.1$~mag, which is plausible.
\item
A comparison of the central star parameters with evolutionary tracks is
fraught with many uncertainties. The progenitor of the central star had a mass
of at least 3~$M_\odot$, but maybe as high as 7~$M_\odot$.
\item
We confirm the finding of \citet{KH96} that the nebula is carbon rich, and we
derive an improved value of the C/O ratio of 2.1. The nebular abundances are
typical for a type IIa PN. They show a moderately high N/O ratio and a mildly
enhanced He abundance.
\item
With the photoionization model we determined that the grains in the ionized
nebula are large. If we use single-sized grains, they would have to have a
radius of 0.15~$\mu$m to explain the far-IR SED. Most likely these large
grains were inherited from the AGB phase.
\item
We use the photoionization model to calculate synthetic photometry in various
mid- and far-IR photometric systems. That way we can determine how much line
and continuum emission are contributing in each band. This analysis enabled us
to determine that there is a continuum component contributing around 25~$\mu$m
that is not accounted for by the Cloudy model. This component is however
detected in the IRS spectrum. The most plausible explanation is that this
comes from stochastically heated very small grains (possibly PAHs). These can
survive in dense H$_2$ emitting clumps in the torus (together with the larger
grains) where they are shielded from the ionizing radiation. The presumed
presence of very small grains could also explain the low $R_V$ value we find
in our dereddening law. However, we know of no direct evidence for the
existence of very small grains and more research will be needed to investigate
this point.
\item
The PACS 70/160~$\mu$m temperature map shows evidence for two radiation
components heating the grains. One components is confined to the region inside
the torus and is quickly absorbed away further out. The other component shows
no evidence for extinction and produces roughly spherical isothermal contours.
Using the Cloudy model we could identify the first component as direct
emission from the central star, while the second component is diffuse emission
from the ionized gas (mainly Ly$\alpha$).
\item
It has previously been suggested by \citet{Hora04} and \citet{RL08} (based on
IRAC images) that NGC~650 contains a PDR surrounding the ionized region. We
show that this assumption is incorrect. All the neutral material (including
H$_2$) in this nebula is contained in dense clumps inside the ionized region.
These clumps have noticeable internal extinction. This is unusual for such a
highly evolved PN.
\end{itemize}

\acknowledgements

We kindly thank Gerardo Ramos-Larios for making the H$\beta$ image of NGC~650
available to us. We thank Nicholas Lee for his invaluable help in interpreting
the Chandra data and we thank the referee for his helpful comments.
PvH and the PACS ICC in Leuven wish to acknowledge support from the Belgian
Science Policy office through the ESA PRODEX programme.
PACS has been developed by a consortium of institutes led by MPE (Germany) and
including UVIE (Austria); KU Leuven, CSL, IMEC (Belgium); CEA, LAM (France);
MPIA (Germany); INAFIFSI/ OAA/OAP/OAT, LENS, SISSA (Italy); IAC (Spain). This
development has been supported by the funding agencies BMVIT (Austria),
ESA-PRODEX (Belgium), CEA/CNES (France), DLR (Germany), ASI/INAF (Italy), and
CICYT/MCYT (Spain).
SPIRE has been developed by a consortium of institutes led by Cardiff Univ.
(UK) and including: Univ. Lethbridge (Canada); NAOC (China); CEA, LAM
(France); IFSI, Univ. Padua (Italy); IAC (Spain); Stockholm Observatory
(Sweden); Imperial College London, RAL, UCL-MSSL, UKATC, Univ. Sussex (UK);
and Caltech, JPL, NHSC, Univ. Colorado (USA). This development has been
supported by national funding agencies: CSA (Canada); NAOC (China); CEA, CNES,
CNRS (France); ASI (Italy); MCINN (Spain); SNSB (Sweden); STFC, UKSA (UK); and
NASA (USA).
Some of the data presented in this paper were obtained from the Multimission
Archive at the Space Telescope Science Institute (MAST). STScI is operated by
the Association of Universities for Research in Astronomy, Inc., under NASA
contract NAS5-26555. Support for MAST for non-HST data is provided by the NASA
Office of Space Science via grant NNX09AF08G and by other grants and
contracts.
The IRS was a collaborative venture between Cornell University and Ball
Aerospace Corporation funded by NASA through the Jet Propulsion Laboratory and
Ames Research Center.
This publication makes use of data products from the Wide-field Infrared
Survey Explorer, which is a joint project of the University of California, Los
Angeles, and the Jet Propulsion Laboratory/California Institute of Technology,
funded by the National Aeronautics and Space Administration.
Data presented in this paper were analysed using ``HIPE'', a joint
development by the Herschel Science Ground Segment Consortium, consisting of
ESA, the NASA Herschel Science Center, and the HIFI, PACS and SPIRE consortia.
This research made use of tools provided by Astrometry.net.
This research made use of the Atomic Line List available at
http://www.pa.uky.edu/\textasciitilde peter/atomic.

\bibliographystyle{aa}
\bibliography{n650}

\end{document}